\documentclass[12pt,preprint]{aastex}

\shorttitle{K-band Spectroscopy of Seyfert 2 Nuclei}
\shortauthors{Imanishi and Alonso-Herrero}

\begin{document}

\title{Near-infrared K-band Spectroscopic Investigation of Seyfert 2 Nuclei 
in the CfA and 12$\mu$m Samples}

\author{Masatoshi Imanishi\altaffilmark{1}}
\affil{National Astronomical Observatory, 2-21-1, Osawa, Mitaka, Tokyo
181-8588, Japan} 
\email{imanishi@optik.mtk.nao.ac.jp} 

\and 

\author{Almudena Alonso-Herrero}
\affil{Departamento de Astrof\'{\i}sica Molecular e Infrarroja, IEM, 
Consejo Superior de Investigaciones Cient\'{\i}ficas, Serrano 113b, 28006
Madrid, Spain} 
\email{aalonso@damir.iem.csic.es}

\altaffiltext{1}{Visiting Astronomer at the Infrared Telescope Facility,
which is operated by the University of Hawaii under
Cooperative Agreement no. NCC 5-538 with the National
Aeronautics and Space Administration, Office of Space
Science, Planetary Astronomy Program.}

\begin{abstract}
We present near-infrared $K$-band slit spectra of the
nuclei of 25 Seyfert 2 galaxies in the CfA and 12 $\mu$m samples.   
The strength of the CO absorption features at 2.3--2.4 $\mu$m 
produced by stars is measured in terms of a spectroscopic CO index.  
A clear anti-correlation between the observed CO index and 
the nuclear $K-L$ color is present, suggesting that a featureless hot
dust continuum heated by an AGN contributes significantly to the
observed $K$-band fluxes in the nuclei of Seyfert 2 galaxies.  
After correction for this AGN contribution, we estimate nuclear stellar
$K$-band luminosities for all sources, and CO indices for sources with
modestly large observed CO indices. 
The corrected CO indices for 10 (=40\%) Seyfert 2 nuclei are found to be
as high as those observed in star-forming or elliptical (=spheroidal)
galaxies. 
We combine the $K$-band data with  measurements of the $L$-band 3.3
$\mu$m polycyclic aromatic hydrocarbon (PAH) emission feature, another
powerful indicator for star-formation, and 
find that the 3.3 $\mu$m PAH to $K$-band stellar luminosity ratios are
substantially smaller than those of starburst galaxies.   
Our results suggest that the 3.3 $\mu$m PAH emission originates in
the putative nuclear starbursts in the dusty tori surrounding the AGNs,
because of its high surface brightness, whereas the $K$-band CO
absorption features detected at the nuclei are dominated by old bulge
(=spheroid) stars, and thus may not be a powerful indicator for the
nuclear starbursts. 
We see no clear difference in the strength of the CO absorption and PAH
emission features between the CfA and 12 $\mu$m Seyfert 2s.
\end{abstract}


\section{Introduction}

According to the current unification paradigm of Seyfert galaxies ---the
most numerous class of Active Galactic Nuclei (AGNs) in the local
universe---, the two types of Seyfert galaxies, type 1 (which show broad
optical emission lines) and type 2 (which do not), are both powered by a
central mass-accreting supermassive black hole. 
The different emission properties are explained by the presence of 
a torus-shaped structure of molecular gas and dust surrounding the central
supermassive black hole (the so-called ``dusty torus'') and the different
viewing angles toward the torus \citep{ant93}.  
Since the dusty torus is rich in molecular gas, it is a natural site for 
starbursts to occur \citep{fab98}. 
Such a nuclear starburst in the torus could play an important role for
AGN physics \citep{wad02}, so that its observational understanding is
very important.  

Attempts to put observational constraints on the nuclear 
starbursts in the torus in Seyfert 2 galaxies have been made extensively,
from UV--optical (e.g., Gonzalez Delgado et al. 1998), through infrared 
(e.g., Oliva et al. 1999; Ivanov et al. 2000; Imanishi 2002; 2003), to
millimeter wavelength ranges \citep{koh02}.    
Among these methods, infrared diagnostics have two major advantages.
First, since nuclear starbursts in the torus are expected to have some
amount of dust extinction, infrared observations are better suited to
investigate their properties than those in the UV or optical, 
because of the reduced effects of dust extinction in the infrared.  
Second, compared to  millimeter observations, a systematic study of 
the nuclear starbursts in a statistically meaningful number of Seyfert
2s is possible with a reasonable amount of telescope time. 
\citet{ima03} obtained infrared $L$-band (2.8--4.1 $\mu$m) slit
spectra of 32 Seyfert 2s in the CfA \citep{huc92} and 12 $\mu$m
(Rush, Malkan, \& Spinoglio 1993) samples.   
Utilizing the polycyclic aromatic hydrocarbon (PAH) emission feature 
at 3.3 $\mu$m as an excellent tool to disentangle star-formation 
from AGN activity \citep{ima02}, the author discussed the
properties of nuclear starbursts in Seyfert 2s in a quantitative manner. 

Nuclear starbursts in Seyfert 2s can also be investigated  through 
infrared $K$-band (2.0--2.5 $\mu$m) spectroscopy. 
Since the CO absorption features  at $\lambda_{\rm rest}$ = 2.3--2.4
$\mu$m in the rest-frame are produced by stars, and not by a pure AGN,
these features can be a good tool to distinguish between stars and an AGN
\citep{oli99,iva00}.   
Among the 32 Seyfert 2s studied by \citet{ima03}, 
25 sources were observed with SpeX at the IRTF 3m telescope 
\citep{ray03} (see Table 2 of Imanishi 2003). 
Since Spex enables to obtain $K$- and $L$-band spectra simultaneously,
the stellar emission in Seyfert 2 nuclei can be investigated also from 
the $K$-band CO absorption features, independently from the
$L$-band PAH diagnostic.  
In this paper, we present $K$-band spectroscopic results and combine
them with the $L$-band results to provide a better understanding of
Seyfert 2 nuclei.    
$H_{0}$ $=$ 75 km s$^{-1}$ Mpc$^{-1}$, $\Omega_{\rm M}$ = 0.3, and
$\Omega_{\rm \Lambda}$ = 0.7 are adopted throughout this paper. 

\section{Targets, Observations and Data Analysis}

Detailed information and the observing log of the 25 Seyfert 2s are
summarized in Table~\ref{tbl-1}.  
The original sample selection criteria and observing details are found
in \citet{ima03}.  
In short, Seyfert 2s in the CfA and 12 $\mu$m samples at $z$ =
0.008--0.035 and at declination $>$ $-$35$^{\circ}$ were selected.
The CfA and 12 $\mu$m Seyfert 2 galaxies were selected based on their
host galaxy magnitudes in the optical \citep{huc92} and {\it IRAS} 12
$\mu$m fluxes \citep{rus93}, respectively.  
The observations were made with SpeX at the IRTF 3m telescope, using the
1.9--4.2 $\mu$m cross-dispersed mode with a 1$\farcs$6 wide slit
\citep{ray03}, which provides a spectral resolution of R $\sim$ 500 in
the $K$-band. 
The 1$\farcs$6 aperture used corresponds to physical scales of between 
250 pc ($z$ = 0.008) and 1 kpc ($z$ = 0.035), so our spectra are
suitable to investigate the {\it nuclear} starbursts, rather than 
ring-shaped {\it circumnuclear} starbursts which are
found typically at distances of $\sim$1 kpc from the nuclei
\citep{lef01}. 
The sky was clear and the seeing  at $K$ was
0$\farcs$6--0$\farcs$9 (full-width at half-maximum; FWHM) throughout the
observations.   

The data reduction of the $K$-band spectra was made using IRAF 
\footnote{
IRAF is distributed by the National Optical Astronomy Observatories,
which are operated by the Association of Universities for Research
in Astronomy, Inc. (AURA), under cooperative agreement with the
National Science Foundation.} 
in the same way as the $L$-band data, as described by \citet{ima03}.  
The Seyfert 2 and standard star spectra were extracted by integrating
signals over 1$\farcs$8--3$\farcs$0 along the slit, depending on the 
actual signal profiles.  
The $K$-band magnitudes of the standard stars in
Table~\ref{tbl-1} were estimated from their $V$-band (0.6 $\mu$m)
magnitudes, adopting $V-K$ colors appropriate to the stellar types
of individual standard stars \citep{tok00}.
The Seyfert 2 spectra were divided by those of the corresponding standard
stars, and then multiplied by the spectra of blackbodies with
temperatures corresponding to those of the individual standard stars
(Table~\ref{tbl-1}). 
Appropriate spectral binning was employed, particularly for faint
sources.  
The intrinsic CO absorption of the adopted standard stars (A-, F-,
G-type main sequence stars) is too weak \citep{arn89} to affect our main
conclusions about the CO absorption strength of the Seyfert 2 nuclei. 

\section{Results}

Figure~\ref{fig1} shows the flux-calibrated $K$-band slit spectra of the 
25 Seyfert 2s in our sample.  
Some sources showed a larger scatter in the spectra over the
$\lambda_{\rm obs}$ = 2.00--2.02 $\mu$m and 2.05--2.07 $\mu$m range in the
observed frame than their neighboring wavelengths, due to the presence
of the Earth's atmospheric absorption features.  
Since these data points may have bad effects on the continuum
determination, with which to estimate the strength of the CO absorption
features, we only used spectra at  $\lambda_{\rm obs}$ = 2.07--2.5 $\mu$m.
For some of the Seyfert 2 galaxies in our sample, spectra with a 
smaller wavelength coverage, in the close vicinity of the Br$\gamma$ emission
line at $\lambda_{\rm rest}$ = 2.166 $\mu$m, are available
\citep{vei97,rhe00}.  
For all the CfA Seyfert 2s in our sample except NGC 4501 and NGC 5347,
\citet{iva00} presented moderate-resolution (R $\sim$ 700) spectra at 
$\lambda_{\rm rest}$ = 2.0--2.4 $\mu$m and/or higher-resolution 
(R $>$ 1000) spectra in narrower wavelength ranges near the CO
absorption features. 
However, for the majority of the 25 Seyfert 2s, our spectra
in Figure~\ref{fig1} are the first moderate-resolution spectra with 
an almost complete $K$-band coverage. 
Thus, a systematic and consistent investigation of the CO absorption
features in a statistically larger number of Seyfert 2s is possible.    

To investigate the presence of  nuclear starbursts in Seyfert 2s
using the strength of the CO absorption features, we basically follow
the method proposed by Doyon, Joseph, \& Wright (1994), who defined a
spectroscopic CO index as  
\begin{eqnarray}
{\rm CO_{spec}} & \equiv & -2.5\log <R_{2.36}>, 
\end{eqnarray}
where $<$R$_{2.36}$$>$ is the average of actual signals at 
$\lambda_{\rm rest}$ = 2.31--2.40 $\mu$m divided by a power-law
continuum (F$_{\rm \lambda}$ = $\alpha \times \lambda^{\beta}$)
extrapolated from shorter wavelengths.    
When the CO absorption is present, $<$R$_{2.36}$$>$ should be less than
unity and thus the CO$_{\rm spec}$ should show a positive value. 
This CO index is insensitive to spectral resolution, because an
average value within the wavelengths that show CO absorption is used.  
Thus, this index can be directly compared among independent spectra 
observed with slightly different resolutions.  

To determine the continuum level, \citet{doy94} used data points at
$\lambda_{\rm rest}$ = 2.0--2.29 $\mu$m, excluding clear emission lines,  
and extrapolated it to $\lambda_{\rm rest}$ $>$ 2.31 $\mu$m. 
In our sample, the redshifts are in the range 
$z =$ 0.008--0.035 and only data points at wavelengths
$\lambda_{\rm obs}$ = 2.07--2.5 $\mu$m are available. 
Thus, data points at wavelengths as short as $\lambda_{\rm rest}$ = 2.0
$\mu$m are not fully covered.  We thus used data points at 
$\lambda_{\rm rest}$ = 2.1--2.29 $\mu$m to determine the continuum level
and extrapolated it to longer wavelengths. 
Following \citet{doy94}, a power-law shape was assumed for the continuum 
and clear emission lines, such as H$_{2}$ 1--0 S(1) 
($\lambda_{\rm rest}$ = 2.122 $\mu$m) or Br$\gamma$ 
($\lambda_{\rm rest}$ = 2.166 $\mu$m), were excluded. 
Regarding the wavelength range that should contain the CO absorption
features, our spectra of $\lambda_{\rm obs}$ = 2.07--2.5 $\mu$m covers
$\lambda_{\rm rest}$ = 2.31--2.4 $\mu$m for all the observed Seyfert 2s
at $z =$ 0.008--0.035, so that the same wavelength range of 
$\lambda_{\rm rest}$ = 2.31--2.4 $\mu$m as employed by \citet{doy94} was
adopted to measure the CO absorption strength.   
Although the wavelength range used for the continuum determination is
slightly different, our spectroscopic CO index is essentially the same
as that defined by \citet{doy94}.   

Whenever possible, we tried to use this method to determine the
continuum level in a consistent way. 
However, in the case of NGC 262 and F04385$-$0828, the extrapolated
continuum determined in this way provided a negative value for 
CO$_{\rm spec}$, possibly because of the curvature of the continuum
emission at $\lambda_{\rm rest}$ = 2.1--2.29 $\mu$m. 
In these two sources, we used only data at $\lambda_{\rm rest}$ =
2.22--2.29 $\mu$m and 2.18--2.29 $\mu$m, respectively, which provided
more reasonable extrapolated continuum levels at 
$\lambda_{\rm rest}$ = 2.31--2.40 $\mu$m.  
The adopted continuum levels are plotted as solid lines in
Fig.~\ref{fig1}. 

The values and histogram of the observed spectroscopic CO indices 
(CO$_{\rm spec-obs}$) are summarized in column 2 of Table~\ref{tbl-2}
and Figure 2a, respectively.  
The largest uncertainty of the CO$_{\rm spec-obs}$ values comes from the  
ambiguity of the continuum determination, which is difficult to assess
quantitatively.  
For some Seyfert 2s, we slightly changed the wavelength range used for the
continuum determination as well as the removed data points which may be 
contaminated by emission lines. 
The CO$_{\rm spec-obs}$ value differs by a factor of as high as 0.03,
which can be taken as a possible maximum uncertainty for CO$_{\rm spec-obs}$.  

For the majority of the CfA Seyfert 2s, \citet{iva00} estimated 
spectroscopic CO indices  based on independent data.
They used a very narrow wavelength range of $\lambda_{\rm rest}$ =
2.285$\pm$0.005 $\mu$m and 2.298$\pm$0.005 $\mu$m for the continuum and
the data points which include CO absorption features, respectively.   
If this method were to be applied to our lower resolution spectra, 
the resulting
uncertainty for CO$_{\rm spec-obs}$ would be large, because only one or
two data points would be included in each of these wavelength ranges.
In Table~\ref{tbl-2} we compare the observed spectroscopic CO
indices based on our wider wavelength coverage spectra, with the indices
derived by \citet{iva00}.    
For some sources (NGC 5929, NGC 7674, and NGC 7682), both  CO index
measurements are in good agreement, but there is some discrepancy for
other sources, possibly because of the different method to derive
CO$_{\rm spec-obs}$ and/or different aperture size.

In some objects, narrow ($<$500 km s$^{-1}$ in FWHM) emission lines are
discernible in the spectra.  
In Fig.1, we indicate the identification of some clear emission lines. 
In general, their equivalent widths are smaller than in starburst 
galaxies \citep{gol97b,van98,coz01}.  
To quantitatively estimate these weak and narrow emission line fluxes,
by tracing the line profiles, spectra with R $>$ 1000 are usually
required. 
No quantitative estimates for the line fluxes are made based on our
lower resolution (R $\sim$ 500) spectra.  

\section{Discussion}

\subsection{CO Absorption Strength}

Unlike pure star-forming galaxies, a featureless continuum  from
hot dust heated by an AGN can contribute significantly to the observed
$K$-band fluxes in Seyfert 2 galaxies.  
This contribution reduces the depth of the CO absorption feature; the
degree of this reduction can vary, depending on the amount of AGN
emission in the $K$-band.   
Indeed, the CO$_{\rm spec-obs}$ values of a large fraction of the observed
Seyfert 2 galaxies in Fig.2a are substantially smaller than
the typical values observed in star-forming or elliptical (=spheroidal)  
galaxies (CO$_{\rm spec}$ $>$ 0.15; Goldader et al. 1997a,b; Ivanov et
al. 2000). 
This is likely to be caused by dilution of the CO
absorption feature by the AGN's featureless continuum emission.    

In the $L$-band, the emission from stars is much reduced 
when compared to that in the $K$-band, whereas hot ($\sim$1000K) dust
heated by an AGN has strong emission both at $K$ and $L$.
Thus, the relative contribution from the AGN to the nuclear spectra
increases substantially from $K$ to $L$ \citep{alo96,alo01}.
In fact, the typical $K - L$ color of an AGN is usually larger (redder)
than a normal stellar population \citep{wil84,alo03}. 
The $K-L$ colors can be used to estimate the AGN contribution to
the nuclear $K$-band spectra. 
The nuclear $K-L$ colors, derived from our slit spectra, are summarized
in  column 5 of Table 2. Figure 3a compares the nuclear $K-L$ colors with 
the CO$_{\rm spec-obs}$ values.  
The $K-L$ color and CO$_{\rm spec-obs}$ are anti-correlated in such a
way that objects with redder $K-L$ colors tend to show smaller 
CO$_{\rm spec-obs}$.  
The probability that a correlation is not present is found to be 
0.01\% \citep{iso86}.  
Since the observed $K-L$ colors should become redder, when 
compared to those of a normal stellar population, for an increasing
contribution from the AGN, the anti-correlation in Fig.3a suggests that
the smaller CO$_{\rm spec-obs}$ values in objects with redder
$K-L$ colors are caused by the AGN contribution. 
This clearly demonstrates that to investigate the stellar emission 
from the CO absorption features in more detail, a correction for the AGN
dilution is necessary.  
To make this correction quantitatively, we make use of the rest-frame
equivalent width of the 3.3 $\mu$m PAH emission feature (EW$_{\rm 3.3PAH}$).
The EW$_{\rm 3.3PAH}$ values are summarized in  column 6 of Table 2.

Hot dust heated by an AGN produces a featureless $L$-band continuum,
while star-forming galaxies usually show strong 3.3 $\mu$m PAH
emission with EW$_{\rm 3.3PAH}$ $\sim$ 100 nm \citep{imd00}. 
Since the equivalent width is by definition robust to dust extinction, 
we assume that pure star-formation shows EW$_{\rm 3.3PAH}$ = 100 nm,
while a pure AGN shows EW$_{\rm 3.3PAH}$ = 0 nm. 
For the three Seyfert 2s with EW$_{\rm 3.3PAH}$ values $>$40 nm (NGC
7682, Mrk 938, and NGC 3660; Imanishi 2003), 
a significant contribution from star-formation to the $L$-band fluxes 
must be present.
For these sources, we estimate the fractional stellar contribution at
$L$, based on the observed EW$_{\rm 3.3PAH}$ values.
More specifically, if the observed EW$_{\rm 3.3PAH}$ value is 70 nm, we
regard that 70\% of the observed $L$-band flux comes from
star-formation, and the remaining 30\% originates in AGN emission. 
We then extract $L$-band fluxes originated in AGN activity from the
observed $L$-band flux levels.  
For the Seyfert 2s with very small EW$_{\rm 3.3PAH}$ of $\lesssim$ 20 nm
(22 sources except NGC 7682, Mrk 938, and NGC 3660), we regard 
the observed  $L$-band fluxes as dominated by AGN-powered 
emission \citep{ima03}.
Namely, the observed $L$-band fluxes are taken as AGN-originated
$L$-band fluxes.   

To derive the AGN fraction in the $K$-band spectra from the estimated
AGN-powered $L$-band flux levels, we assume that the intrinsic AGN
colors in all Seyfert 2 galaxies are $K-L$ = 2.79, as was done by
\citet{iva00}. 
\citet{alo03} have shown that the infrared AGN 
spectral energy distributions in Seyfert 2 galaxies are fitted with
f$_{\nu}$ $\propto$ $\nu^{-3}$, which provides $K-L$ $\sim$ 2.5--2.6. 
Thus, the adopted assumption of $K-L$ = 2.79 is reasonably justified
from the observational point of view. 
After subtracting this AGN emission from the observed $K$-band spectra,
we can infer the stellar $K$-band emission and thereby 
obtain corrected CO indices (CO$_{\rm spec-cor}$).  
The derived stellar $K$-band luminosities are summarized in column 7 of
Table 2.  
Regarding the corrected CO index (CO$_{\rm spec-cor}$), 
if the AGN contribution to the observed $K$-band flux is estimated to be
60\%, then the CO absorption optical depth increases by 2.5
\{=1/(1$-$0.6)\}, and the spectroscopic CO index increases accordingly. 
Thus, when the observed CO index is very small, say CO$_{\rm spec-obs}$ $<$
0.1, due to dilution from AGN emission at $K$, a small uncertainty in
the continuum determination can result in a large ambiguity 
in the corrected CO index. 
There are 10 sources with CO$_{\rm spec-obs}$ $>$ 0.1. 
We regard the CO absorption as clearly detected in these 10 sources, and
estimate CO$_{\rm spec-cor}$ only for them, in order for the
correction process to work reasonably. 
The CO$_{\rm spec-cor}$ values of the 10 sources are summarized 
in column 3 of Table 2, and the histogram of the CO indices corrected in
this way is plotted in Figure 2b.
Although the AGN correction certainly increases the CO indices, the
change of their overall distribution is little.
Thus, possible ambiguities of this AGN correction do not seriously
affect our main conclusions. 

This method to correct the CO indices for AGN emission using our own
slit spectra has important advantages over some other methods, such as
that based on nuclear $K$- and $L$-band photometric data.
First, since the $K$- and $L$-band spectra were taken simultaneously
under the same observing conditions and probe the same regions in the
observed Seyfert 2 galaxies, there is little ambiguity on 
the flux inter-calibration.
Second, the compact nuclear $L$-band emission observed in Seyferts is usually
assumed to be dominated by AGN emission \citep{iva00}.
Although this is true for the majority of Seyfert 2s
\citep{ima02,ima03}, there are some exceptions.  
From our $L$-band slit spectra, we can clearly pick out these exceptions
based on the EW$_{\rm 3.3PAH}$ values and take into account the stellar
contamination to the nuclear $L$-band fluxes. 

However, there is one caution for this correction process.
The PAH emission observed in star-forming regions primarily comes from
photo-dissociation regions located between HII regions and molecular
gas, and not from the HII regions themselves \citep{sel81}. 
If the star-forming regions have such a high radiation density that
their emitting volumes are predominantly occupied by HII regions with
virtually no photo-dissociation regions, then no 3.3 $\mu$m PAH emission
is expected.   
Although this is the case for the nuclear star cluster in the dwarf
galaxy NGC 5253 \citep{alo04}, this kind of star-forming regions are 
probably minor exceptions.  
Figure 3b compares the nuclear $K - L$ color with 
EW$_{\rm 3.3PAH}$ for our sample of Seyfert 2s. 
Using a statistical test by Isobe et al. (1986), the probability that
a correlation is not present is found to be 30\%, suggesting an
anti-correlation between the $K - L$ and EW$_{\rm 3.3PAH}$,    
as expected from the assumption that the featureless $L$-band continuum
originates in the AGN. 
If emission from pure HII regions were contributing significantly to the
nuclear $L$-band featureless continuum in many Seyfert 2 galaxies, then
the estimated AGN contribution to the nuclear $K$-band flux would be
overestimated.  
In this case, the corrected CO indices would be smaller than the above
estimates.  

In Fig.2b, 40\% (=10/25) of the observed Seyfert 2 galaxies have
corrected CO indices (CO$_{\rm spec-cor}$) that are as large as those of
star-forming or elliptical (=spheroidal) galaxies (CO$_{\rm spec}$ $>$
0.15; Goldader et al. 1997a,b; Ivanov et al. 2000).  
Since very young star-formation, with less than a few million years in age,
should show very small CO$_{\rm spec}$ values \citep{lei99},   
such star-formation is missed in this $K$-band CO absorption method.
Furthermore, when the AGN dilution is very severe in the $K$-band,
the CO absorption features are very weak and so the correction of the 
observed spectroscopic CO indices for the AGN dilution is not possible. 
In fact, for NGC 7674 and F01475$-$0740, although on-going nuclear star
formation is indicated from the detected 3.3 $\mu$m PAH emission
feature, no clear CO absorption features are detected. 
Thus, this fraction (=40\%) should be taken as a lower limit of Seyfert
2 nuclei with stellar emission. 

\subsection{The Nature of the Stellar Emission Detected in the $K$-band}

The stellar signatures detected in the nuclei of Seyfert 2s can be due
either to nuclear starbursts in the AGN torus and/or old bulge
(=spheroid) stars.  
Since both starbursts and old bulge stars show CO$_{\rm spec}$ $\gtrsim$ 0.15, 
these two populations are not easily distinguishable based on the 
CO$_{\rm spec}$ values alone \citep{oli95}. 
To further understand the nature of the detected stellar emission in the
$K$-band, we combine the $K$-band data with the $L$-band spectra 
\citep{ima03}.  
Figure 4 compares observed CO index with EW$_{\rm 3.3PAH}$ (shown in
the column 6 of Table 2). They are weakly correlated.

\citet{ima03} implicitly assumed that the detected 3.3 $\mu$m PAH
emission originates in the putative nuclear starbursts in the AGN 
torus. Here we test this assumption in a quantitative manner based on the
3.3 $\mu$m PAH emission luminosity measured in our apertures, because
starbursts and old bulge stars are distinguishable in terms of emission
surface brightness.   
\citet{hec01} estimated a star-formation rate per unit area of 
10$^{-1}$ M$_{\odot}$ yr$^{-1}$ kpc$^{-2}$ as the
lower limit beyond which the superwind activity, a characteristic 
property of starburts, can occur.  
This limit corresponds to an infrared surface brightness of 
S$_{\rm IR}$ $\sim$ 2 $\times$ 10$^{42}$ ergs s$^{-1}$ kpc$^{-2}$
\citep{ken98}, or a 3.3 $\mu$m PAH emission surface brightness of 
S$_{\rm 3.3PAH}$ $\sim$ 2 $\times$ 10$^{39}$ ergs s$^{-1}$ kpc$^{-2}$ 
\citep{ima02}.
Among the 25 Seyfert 2s studied in this paper, the 3.3 $\mu$m PAH
emission was detected by \citet{ima03} in the seven sources; NGC 7674,
NGC 7682, Mrk 938, F01475$-$0740, F04385$-$0828, NGC 3660, and NGC 4968.
The surface brightnesses of the 3.3 $\mu$m PAH emission are summarized in
the last column of Table 2.
For all the PAH-detected sources, the observed surface brightnesses 
significantly exceed the threshold required for starbursts. 
Therefore, for the PAH-detected Seyfert 2 nuclei, we can conclude that
the detected 3.3 $\mu$m PAH emission must come from a nuclear
starburst, and cannot be of old stellar origin.  

Figure 5 compares the $K$-band stellar luminosity with the 3.3 $\mu$m
PAH emission luminosity from the Seyfert 2 nuclei measured from our
slit spectra.  
In infrared-luminous starburst galaxies, the $K$-band to infrared
luminosity ratios are found to be L$_{\rm K}$/L$_{\rm IR}$ $\sim$
10$^{-1.6}$ \citep{gol97a}.  
The 3.3 $\mu$m PAH to infrared luminosity ratios in starbursts are
estimated to be L$_{\rm 3.3PAH}$/L$_{\rm IR}$ $\sim$ 10$^{-3}$
\citep{mou90,sat95,ima02}.   
Thus, the 3.3 $\mu$m PAH to $K$-band luminosity ratios are expected to
be L$_{\rm 3.3PAH}$/L$_{\rm K}$ $\sim$ 10$^{-1.4}$, if both luminosities
trace starbursts.  
This value is shown as a solid line in Figure 5.
The observed ratios in the Seyfert 2s 
in our sample are smaller, by a large factor, than
the value expected for starbursts. 
If the starbursts are very young, with ages of less than a few million
years, then L$_{\rm K}$/L$_{\rm IR}$ decreases \citep{gol97a}, and thus
the discrepancy between the observed and expected  
L$_{\rm 3.3PAH}$/L$_{\rm K}$ ratios is even larger. 

Figure 5 suggests that in the nuclei of Seyfert 2s, 
compared to starbursts, L$_{\rm 3.3PAH}$ is
underluminous or L$_{\rm K}$ is overluminous.  
One possibility for the L$_{\rm 3.3PAH}$ underluminosity is the
destruction of PAH emitting material by energetic radiation from the
central AGN \citep{voi92}. 
However, since the putative nuclear starbursts are expected to occur at
the outer part of the obscuring torus where the AGN radiation is
significantly attenuated \citep{hec97,ima03}, this destruction is
unlikely to be severe. 
Aside from the PAH destruction by the AGN radiation, the 
PAH emission may be
suppressed if the stellar radiation density is so high that the volume
fraction of photo-dissociation regions relative to HII regions is
substantially smaller than typical starbursts.   
If the PAH destruction in the nuclear starbursts in the
AGN tori were more severe than typical starburst galaxies, the
actual infrared luminosity of the nuclear starbursts would be
substantially larger than the estimate based on the assumption of
L$_{\rm 3.3PAH}$/L$_{\rm IR}$ $\sim$ 10$^{-3}$ \citep{ima03}.  
However, we regard this possibility as unlikely, because 
for a few selected Seyfert 2s for which the luminosities of the
nuclear starbursts were estimated both from the UV and 3.3 $\mu$m PAH 
emission (assuming L$_{\rm 3.3PAH}$/L$_{\rm IR}$ $\sim$ 10$^{-3}$), 
the estimated nuclear starburst luminosities agree quantitatively well
\citep{ima02}.  

We suggest that our $K$- and $L$-band results are more naturally
explained by a scenario where the 3.3 $\mu$m PAH emission
originates in the spatially-unresolved (smaller than sub-arcsec) nuclear
starbursts in the AGN torus, whereas the signatures of stellar emission
detected in the $K$-band spectra are produced mainly by old bulge stars. 
In fact, the nuclear $L$-band emission of Seyfert 2s is dominated by 
spatially-unresolved emission rather than the bulge component,
whereas the extended stellar components often contribute predominantly to
the nuclear $K$-band emission \citep{alo96,alo98,alo01}.  
Old bulge stars can contribute significantly to the nuclear $K$-band
fluxes, but cannot to the PAH emission due to the paucity of
PAH-exciting far-UV photons in bulges compared to starbursts.    
This scenario explains the small L$_{\rm PAH}$/L$_{\rm K}$ ratios
observed in Figure 5.    
In this scenario, the CO absorption strength is primarily due to
old bulge stars, rather than nuclear starbursts, and thus it may not be
a useful tool to detect the putative nuclear starbursts in the AGN torus.  
This scenario is consistent with the conclusion by \citet{iva00}.

\subsection{Do the CfA and 12 $\mu$m Seyfert 2 Galaxies Contain Different
Nuclear Stellar Properties?} 

To improve the statistics, so far we have discussed near-infrared emission
properties by combining both the CfA and 12 $\mu$m Seyfert 2s together. 
However, since selection criteria of these samples are different, 
a separate discussion of Seyfert 2s in each sample is of interest.

Figure 6 plots the distribution of CO$_{\rm spec-obs}$, 
EW$_{\rm 3.3PAH}$, L$_{\rm 3.3PAH}$/L$_{\rm IR}$, and  
L$_{\rm 3.3PAH}$/L$_{\rm K stellar}$. 
In all the plots, no systematic differences are found between 
Seyfert 2s in the CfA and 12 $\mu$m samples. 
Thus, the main arguments in this paper are taken to be applicable to
Seyfert 2 stellar populations in general.
In particular, the L$_{\rm 3.3PAH}$/L$_{\rm IR}$ ratios in Fig.6c and 
L$_{\rm 3.3PAH}$/L$_{\rm K stellar}$ in Fig.6d trace the relative
contribution from the nuclear starbursts to the infrared and nuclear
stellar $K$-band luminosity, respectively \citep{ima03}.
There was a possibility that the 12 $\mu$m Seyfert 2s may be biased to
those with powerful nuclear starbursts \citep{ho01}, because the
starbursts can be strong emitters at 12$\mu$m.  
The absence of systematic differences in Fig.6c and 6d suggests that no
such strong bias is present in the 12 $\mu$m Seyfert 2 sample.

\section{Conclusion}

We have presented results from a near-infrared $K$-band slit
spectroscopic survey of 25 Seyfert 2 galaxies in the CfA and 12 $\mu$m
samples. These data have been combined with results from the $3.3\,\mu$m
PAH emission survey of Imanishi (2003). 
The CO absorption features at $\lambda_{\rm rest}$ = 2.3--2.4 $\mu$m 
have been used to investigate the properties of the stellar
populations in the nuclei of Seyfert 2 galaxies. We have found 
that the observed values of the CO indices in a large 
fraction of Seyfert 2 nuclei are affected by dilution produced by the
featureless continuum associated with the AGN, because Seyfert 2 nuclei
with redder $K-L$ colors tend to show shallower CO absorption features.  
For sources with modestly large observed CO indices ($>$0.1), 
we have corrected the CO indices for AGN dilution.
This correction has made a small change for the overall distribution of
the CO indices in our sample and we have estimated that $\sim$40\%
(=10/25) of Seyfert 2 nuclei have CO indices similar to those of
star-forming or elliptical (=spheroidal) galaxies.  
This is only a lower limit of Seyfert 2 nuclei with stellar emission
because young star-formation shows no CO absorption features and because
in Seyfert 2 nuclei where the $K$-band light is dominated by AGN
emission, the correction of the observed CO indices for AGN dilution is
not possible.  

A comparison between the nuclear 3.3 $\mu$m PAH to stellar $K$-band
luminosity ratio in the nuclei of Seyfert 2s and observations for
infrared luminous starburst galaxies implies that the CO absorption
features detected in the nuclear $K$-band spectra of Seyfert 2 galaxies
are due mainly to old bulge (=spheroid) stars rather than the putative
nuclear starbursts in the AGN torus, and thus that the putative
starbursts are difficult to detect with $K$-band spectra over the scales
probed by our data (250\,pc to 1\,kpc).  
We found no clear difference in the properties of the  2.3 $\mu$m CO
absorption and 3.3 $\mu$m PAH emission features between the CfA and 12
$\mu$m Seyfert 2 samples. 

\acknowledgments

We are grateful to J. Rayner, S. Bus, P. Sears, W. Golisch for
their support during our IRTF SpeX observing runs, to K. Wada for
useful discussions, and to the referee for valuable comments.  
AAH acknowledges support from the Spanish Programa Nacional de
Astronom\'{\i}a y Astrof\'{\i}sica under grant AYA2002-01055.
Some part of the data analysis was made using a computer system operated
by the Astronomical Data Analysis Center (ADAC) and the Subaru Telescope
of the National Astronomical Observatory, Japan.
This research has made use of the SIMBAD database, operated at CDS,
Strasbourg, France, and of the NASA/IPAC Extragalactic Database 
(NED) which is operated by the Jet Propulsion Laboratory, California
Institute of Technology, under contract with the National Aeronautics
and Space Administration.

\clearpage

\begin{deluxetable}{lccccccccc}
\tabletypesize{\scriptsize}
\rotate
\tablecaption{Details of the Observed Seyfert 2 Galaxies \label{tbl-1}}
\tablewidth{0pt}
\tablehead{
\colhead{} & \colhead{} & \colhead{Physical scale} & \colhead{Date} &
\colhead{Integration} & 
\multicolumn{4}{c}{Standard Star}& \colhead{} \\
\cline{6-9} \\
\colhead{Object} & \colhead{Redshift} & \colhead{[kpc]} & \colhead{[UT]} &
\colhead{[min]} & \colhead{Name} & \colhead{K-mag} & \colhead{Type} &
\colhead{T$_{\rm eff}$ (K)} & \colhead{Remark}  \\
\colhead{(1)} & \colhead{(2)} & \colhead{(3)} & \colhead{(4)} & 
\colhead{(5)} & \colhead{(6)} & \colhead{(7)} & \colhead{(8)} &
\colhead{(9)} & \colhead{(10)}  
}
\startdata
Mrk 993  & 0.015 & 0.46 & 2002 Aug 27, 29 & 100 & HR 410 & 5.0 & F7V &
6240 & CfA \\  
Mrk 573  & 0.017 & 0.52 & 2002 Aug 28 & 30 & HR 650  & 4.2 & F8V & 6000
& CfA \\ 
NGC 4388 & 0.008 & 0.25 & 2003 Mar 18 & 30 & HR 4708 & 5.0 & F8V & 6000 & CfA,
12$\mu$m \\ 
NGC 4501 & 0.008 & 0.25 & 2003 Mar 18 & 40 & HR 4708 & 5.0 &
F8V & 6000 & CfA\tablenotemark{a}, 12$\mu$m \\
NGC 5252 & 0.023 & 0.69 & 2003 Mar 19 & 40 & HR 5011 & 3.8 & G0V & 5930
& CfA \\ 
NGC 5347 & 0.008 & 0.25 & 2003 Mar 18 & 40 & HR 5346 & 4.9 & F8V & 6000 & CfA,
12$\mu$m \\ 
NGC 5695 & 0.014 & 0.43 & 2003 Mar 20 & 60 & HR 5630 & 5.0 & F8V & 6000
& CfA \\ 
NGC 5929 & 0.008 & 0.25 & 2002 Aug 29 & 40 & HR 5728 & 4.6 & G3V & 5800 & CfA,
12$\mu$m \\ 
NGC 7674 & 0.029 & 0.87 & 2002 Aug 27 & 20 & HR 8653 & 4.7 & G8IV & 5400 & CfA,
12$\mu$m \\ 
NGC 7682 & 0.017 & 0.52 & 2002 Aug 27, 29 & 160 & HR 8969 & 2.8 & F7V & 6240 &
CfA \\   
\hline
Mrk 938  & 0.019 & 0.58 & 2002 Aug 28 & 30 & HR 8917\tablenotemark{b} & 5.0 &
G0V & 5930 & 12$\mu$m\\ 
NGC 262 (Mrk 348) & 0.015 & 0.46 & 2002 Aug 28 & 30 & HR 410 & 5.0 & F7V & 6240
& 12$\mu$m\\
NGC 513  & 0.020 & 0.61 & 2002 Aug 28 & 30 & HR 410 & 5.0 & F7V & 6240 &
12$\mu$m\\ 
F01475$-$0740 & 0.017 & 0.52 & 2002 Aug 27 & 30 & HR 466 & 4.9 & F7V & 6240 &
12$\mu$m\\  
NGC 1194 & 0.013 & 0.40 & 2002 Aug 28 & 30 & HR 996 & 3.3 & G5V & 5700 &
12$\mu$m\\ 
NGC 1241 & 0.014 & 0.43 & 2002 Aug 29 & 50 & HR 784 & 4.6 & F6V & 6400 &
12$\mu$m\\ 
NGC 1320 (Mrk 607) & 0.010 & 0.31 & 2002 Aug 29 & 30 & HR 784 & 4.6 & F6V &
6400 & 12$\mu$m\\
F04385$-$0828 & 0.015 & 0.46 & 2002 Aug 29 & 30 & HR 1536 & 4.4 & F8V & 6000 &
12$\mu$m\\  
NGC 1667 & 0.015 & 0.46 & 2003 Mar 20 & 60 & HR 1536 & 4.4 & F8V & 6000
& 12$\mu$m \\ 
NGC 3660 & 0.012 & 0.37 & 2003 Mar 20 & 90 & HR 4529 & 4.9 & F7V & 6240
& 12$\mu$m \\ 
NGC 4968 & 0.010 & 0.31 & 2003 Mar 20 & 30 & HR 4935 & 4.3 & F7V & 6240
& 12$\mu$m\\ 
MCG-3-34-64 & 0.017 & 0.52 & 2003 Mar 19 & 40 & HR 4995 & 3.7 & G6V & 5620 &
12$\mu$m\\  
MCG-2-40-4 & 0.024 & 0.72 & 2003 Mar 19 & 40 & HR 5779 & 5.2 & F7V &
6240 & 12$\mu$m\\ 
F15480$-$0344 & 0.030 & 0.90 & 2002 Aug 29 & 30 & HR 5779 & 5.2 & F7V & 6240 &
12$\mu$m\\  
MCG-3-58-7 & 0.032 & 0.96 & 2002 Aug 28 & 20 & HR 8457 & 4.9 & F6V &
6400 & 12$\mu$m\\ 
\enddata

\tablecomments{
Column (1): Object. 
Column (2): Redshift.
Column (3): Physical scale of the employed slit width of 1$\farcs$6.
Column (4): Observing date in UT. 
Column (5): Net on-source integration time in min.
Column (6): Standard star name.
Column (7): Adopted $K$-band magnitude for the standard star. 
Column (8): Stellar spectral type.
Column (9): Effective temperature in K. 
Column (10): CfA \citep{huc92} or 12 $\mu$m \citep{rus93} Seyfert 2s.
}

\tablenotetext{a}{NGC 4501 is classified as a LINER in the CfA sample
\citep{huc92}.} 

\tablenotetext{a}{Flux calibration was made using HR 8457.}

\end{deluxetable}

\clearpage
\begin{deluxetable}{lrrlrrrrr}
\tabletypesize{\scriptsize}
\rotate
\tablecaption{Near-infrared Properties of the Observed Seyfert 2
Galaxies \label{tbl-2}} 
\tablewidth{0pt}
\tablehead{
\colhead{Object} & \colhead{CO$_{\rm spec-obs}$} & 
\colhead{CO$_{\rm spec-cor}$} & 
\colhead{Remarks} & \colhead{K $-$ L} & \colhead{EW$_{\rm 3.3PAH}$} & 
\colhead{L$_{\rm K}$}  & \colhead{L$_{\rm 3.3PAH}$} & 
\colhead{S$_{\rm 3.3PAH}$}  \\ 
\colhead{} & \colhead{} & \colhead{} & \colhead{} &
\colhead{(mag)} & \colhead{(nm)} & 
\colhead{($\times$ 10$^{42}$ ergs s$^{-1}$)} & 
\colhead{($\times$ 10$^{39}$ ergs s$^{-1}$)} & 
\colhead{($\times$ 10$^{39}$ ergs s$^{-1}$ kpc$^{-2}$)} \\ 
\colhead{(1)} & \colhead{(2)} & \colhead{(3)} & \colhead{(4)} & 
\colhead{(5)} & \colhead{(6)} & \colhead{(7)} & \colhead{(8)} & 
\colhead{(9)} 
}
\startdata
Mrk 993  & 0.20 & 0.24 & 0.26 (0.29) & 0.8 & $<$16 & 3.8  & $<$9.4 & $<$30\\ 
Mrk 573  & 0.06 & \nodata & 0.15 (0.19) & 1.8 & $<$6  & 6.0  & $<$18 & $<$50\\
NGC 4388 & 0.03 & \nodata & 0.09        & 2.0 & $<$5  & 1.3  & $<$4.7 & $<$50\\
NGC 4501 & 0.20 & 0.23 & & 0.3 & $<$12 & 3.4  & $<$4.7 & $<$50 \\
NGC 5252 & 0.01 & \nodata & 0.10        & 1.7 & $<$12 & 8.1  & $<$56.4 & $<$80\\
NGC 5347 & 0.03 & \nodata &             & 1.9 & $<$10 & 1.1  & $<$7.9 & $<$90\\
NGC 5695 & 0.22 & 0.25 & 0.16 (0.18) & 0.2 & $<$16 & 3.4  & $<$6.1 & $<$30\\
NGC 5929 & 0.19 & 0.21 & 0.17 (0.21) & 0.4 & $<$13 & 0.97 & $<$1.7 & $<$20\\
NGC 7674 & 0.00 & \nodata & 0.03        & 2.1 & 7     & 17.7 & 136 & 120 \\
NGC 7682 & 0.19 & 0.20 & 0.18 (0.18) & 0.2 & 40    & 3.2  & 11 & 30 \\  
\hline
Mrk 938       & 0.20 & 0.21 & & 0.6 & 75    & 29.1 & 289 & 570 \\
NGC 262       & 0.01 & \nodata & & 2.2 & $<$7  & 4.6  & $<$38 & $<$120\\
NGC 513       & 0.03 & \nodata & & 1.4 & $<$6  & 10.4 & $<$22 & $<$40\\
F01475$-$0740 & 0.00 & \nodata & & 2.0 & 21    & 1.3  & 22 & 50 \\
NGC 1194      & 0.00 & \nodata & & 2.3 & $<$4  & 1.8  & $<$8.8 & $<$40\\
NGC 1241      & 0.25 & 0.29 & & 0.4 & $<$12 & 3.0  & $<$4.1 & $<$20\\
NGC 1320      & 0.08 & \nodata & & 2.0 & $<$8  & 2.0  & $<$12 & $<$90\\
F04385$-$0828 & 0.04 & \nodata & & 2.4 & 4     & 2.1  & 18 & 60 \\
NGC 1667      & 0.07 & \nodata & & 0.5 & $<$16 & 3.3  & $<$7.1 & $<$30\\
NGC 3660      & 0.21 & 0.23 & & 0.7 & 50    & 1.4  & 11 & 50 \\
NGC 4968      & 0.14 & 0.22 & & 1.6 & 18    & 2.1  & 17 & 120 \\
MCG-3-34-64   & 0.00 & \nodata & & 1.9 & $<$6  & 4.9  & $<$18 & $<$50\\
MCG-2-40-4    & 0.04 & \nodata & & 1.7 & $<$3  & 47.1 & $<$80 & $<$110\\
F15480$-$0344 & 0.21 & 0.33 & & 1.5 & $<$11 & 12.9 & $<$59 & $<$50\\
MCG-3-58-7    & 0.06 & \nodata & & 1.9 & $<$3  & 35.0 & $<$78 & $<$60\\
\enddata

\tablecomments{
Column (1): Object. 
Column (2): Observed spectroscopic CO index. 
Column (3): Corrected spectroscopic CO index.
Column (4): Observed (corrected) spectroscopic CO index taken from
\citet{iva00}. 
Column (5): Nuclear $K - L$ color in magnitude, measured with our slit 
spectroscopy.  
Column (6): Rest frame equivalent width of the 3.3 $\mu$m PAH emission
feature in nm, taken from \citet{ima03}.
Column (7): $K$-band stellar luminosity, defined as
$\lambda$L$_{\lambda}$ at 2.2 $\mu$m.
Column (8): Observed 3.3 $\mu$m PAH luminosity, taken from
\citet{ima03}. 
Column (9): Surface brightness of the 3.3 $\mu$m PAH emission feature.
} 

\end{deluxetable}

\clearpage

\begin{figure}
\includegraphics[angle=0,scale=.42]{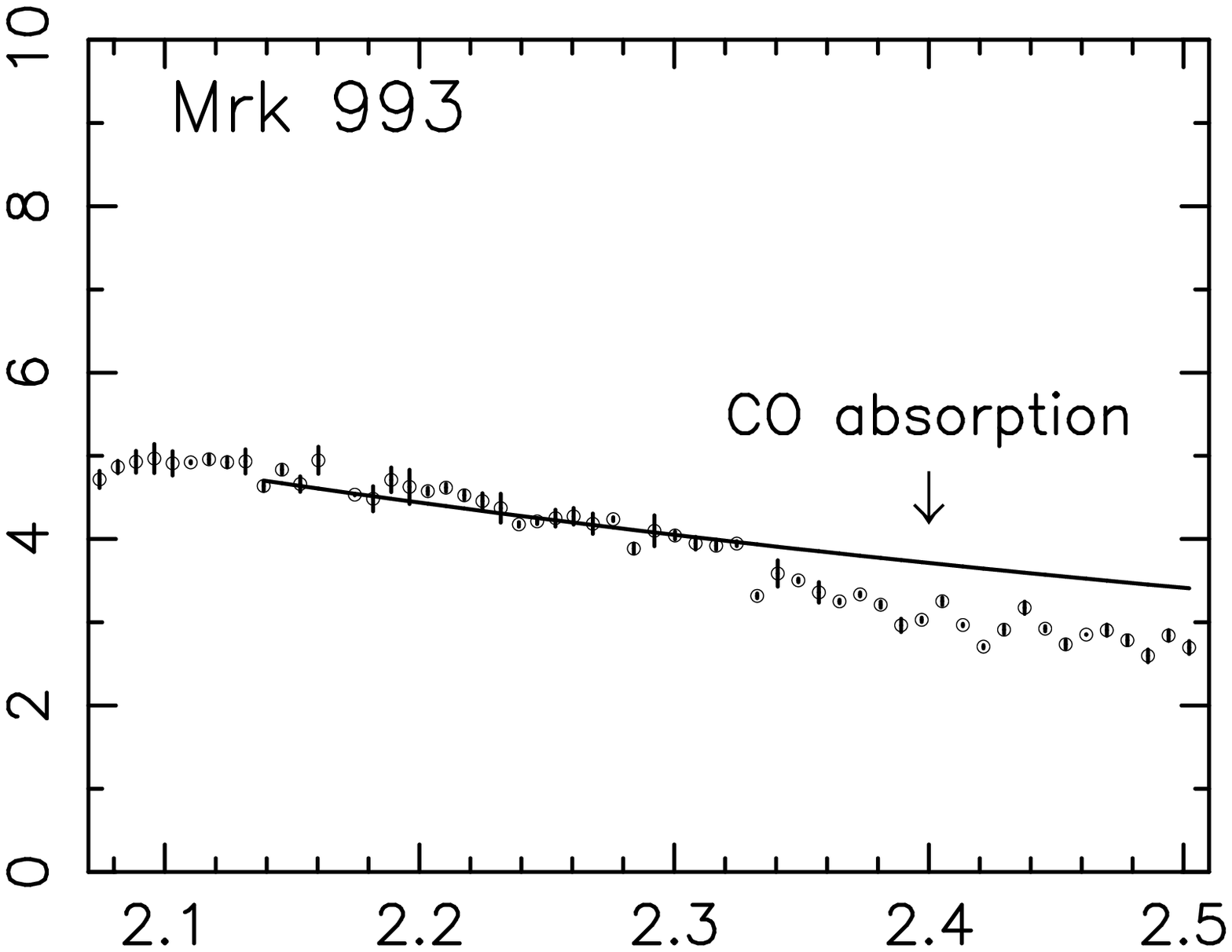} 
\includegraphics[angle=0,scale=.42]{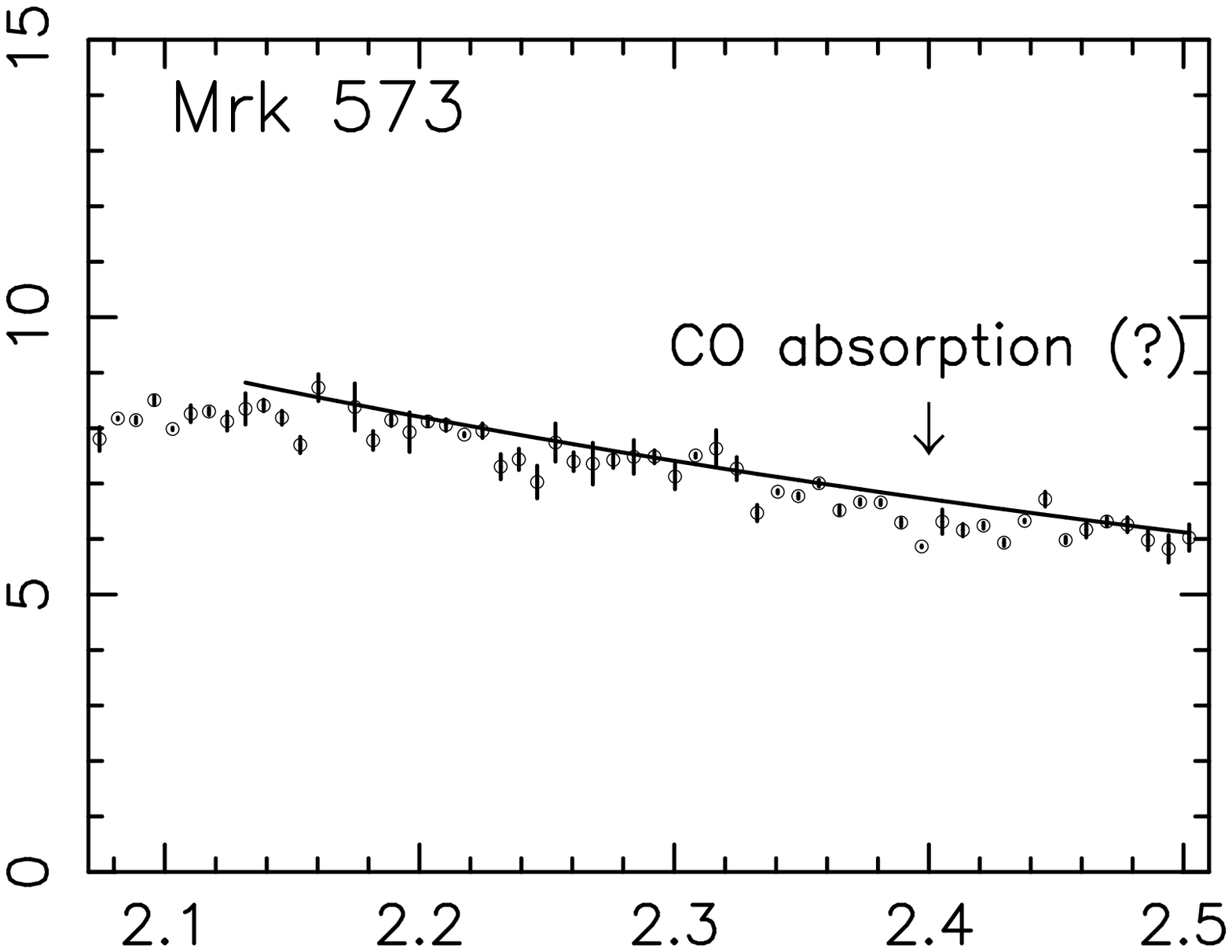} 
\includegraphics[angle=0,scale=.42]{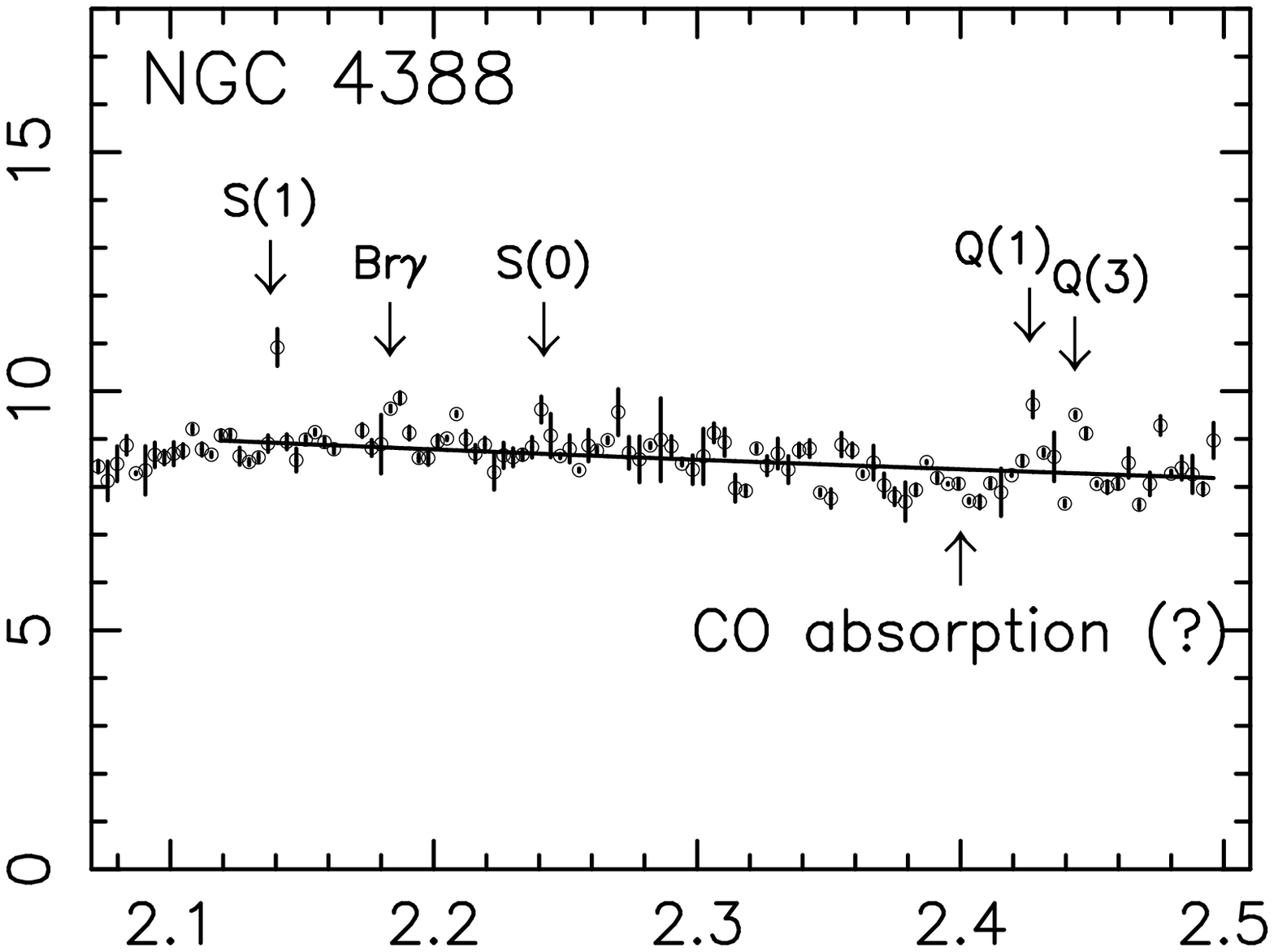} 
\includegraphics[angle=0,scale=.42]{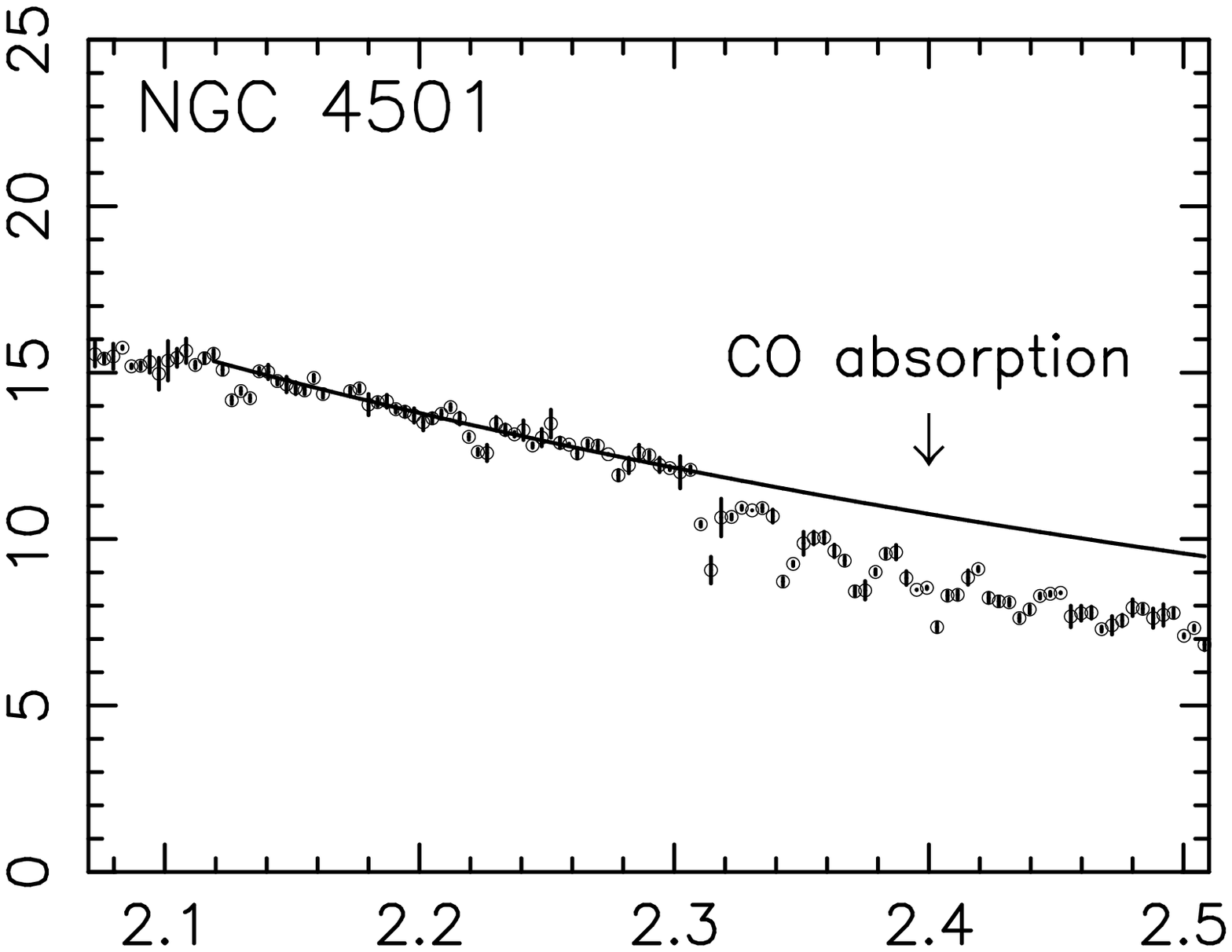} 
\includegraphics[angle=0,scale=.42]{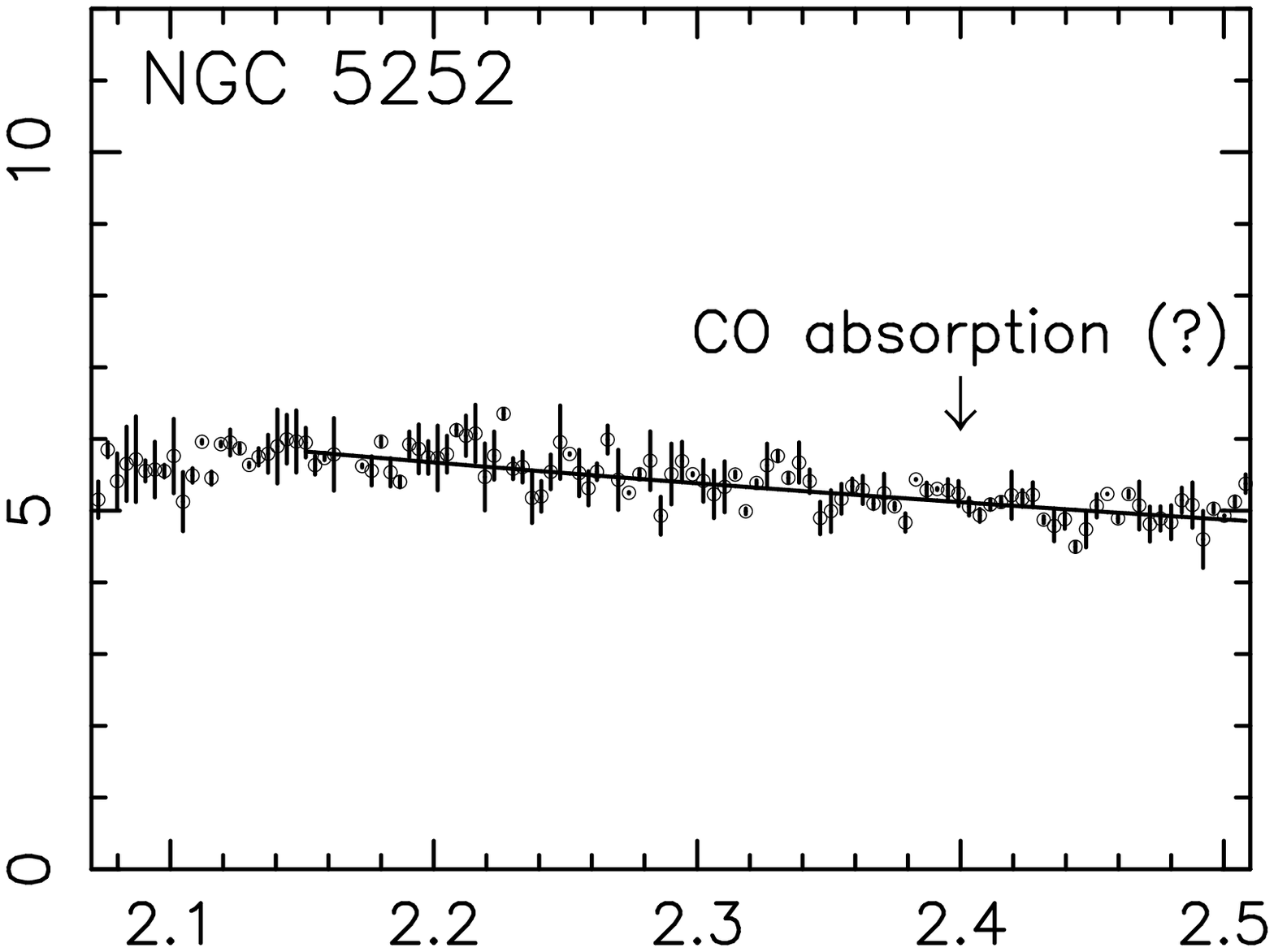} 
\includegraphics[angle=0,scale=.42]{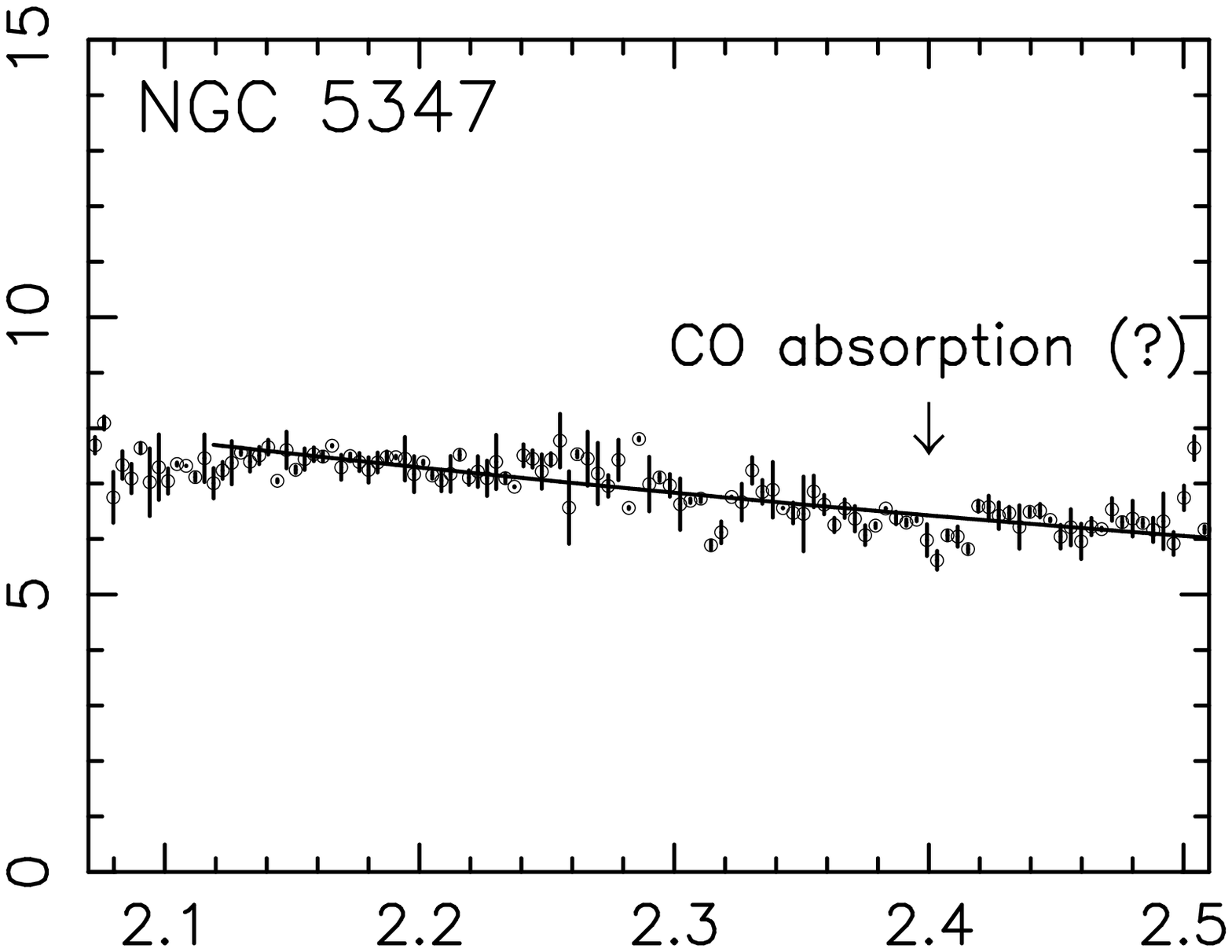} 
\end{figure}

\clearpage

\begin{figure} 
\includegraphics[angle=0,scale=.42]{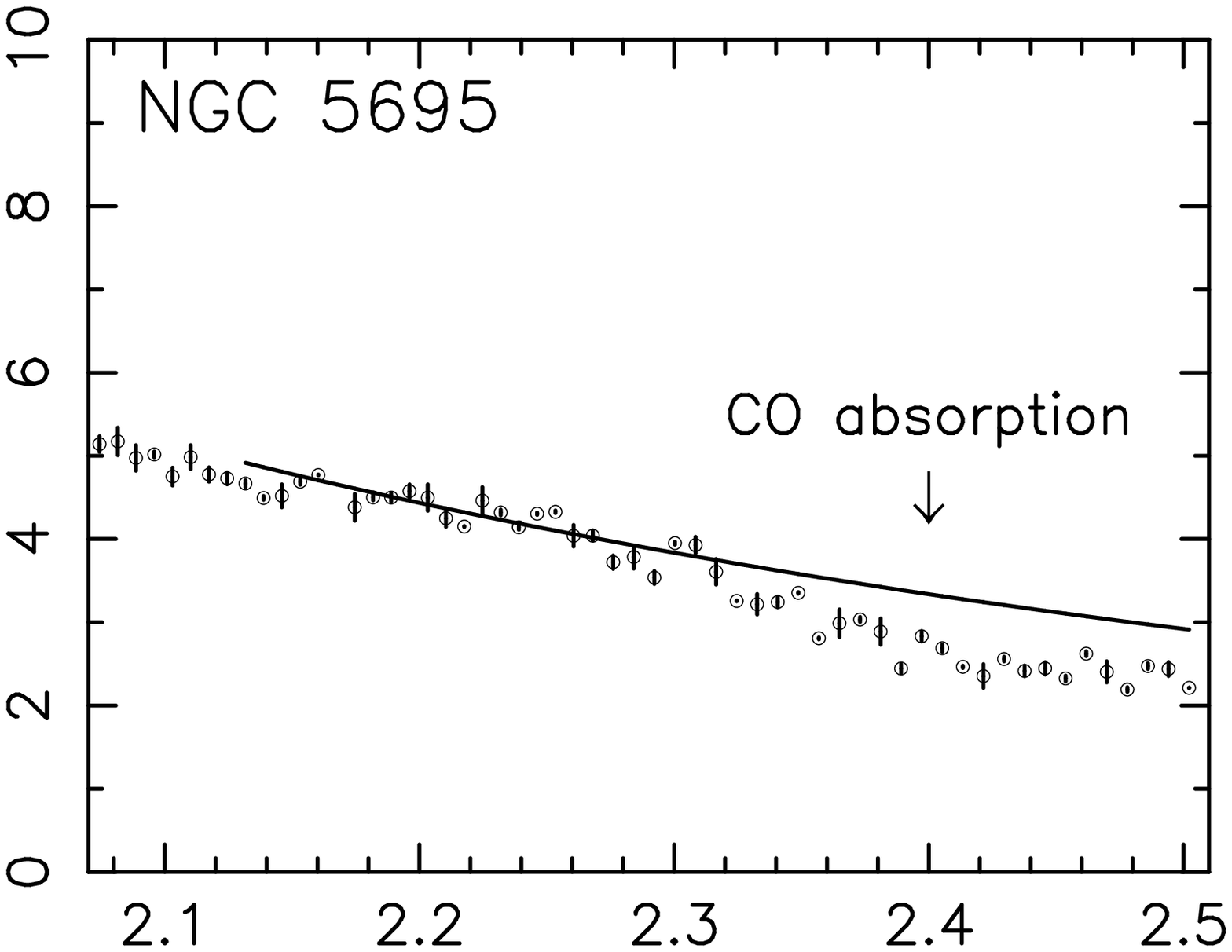} 
\includegraphics[angle=0,scale=.42]{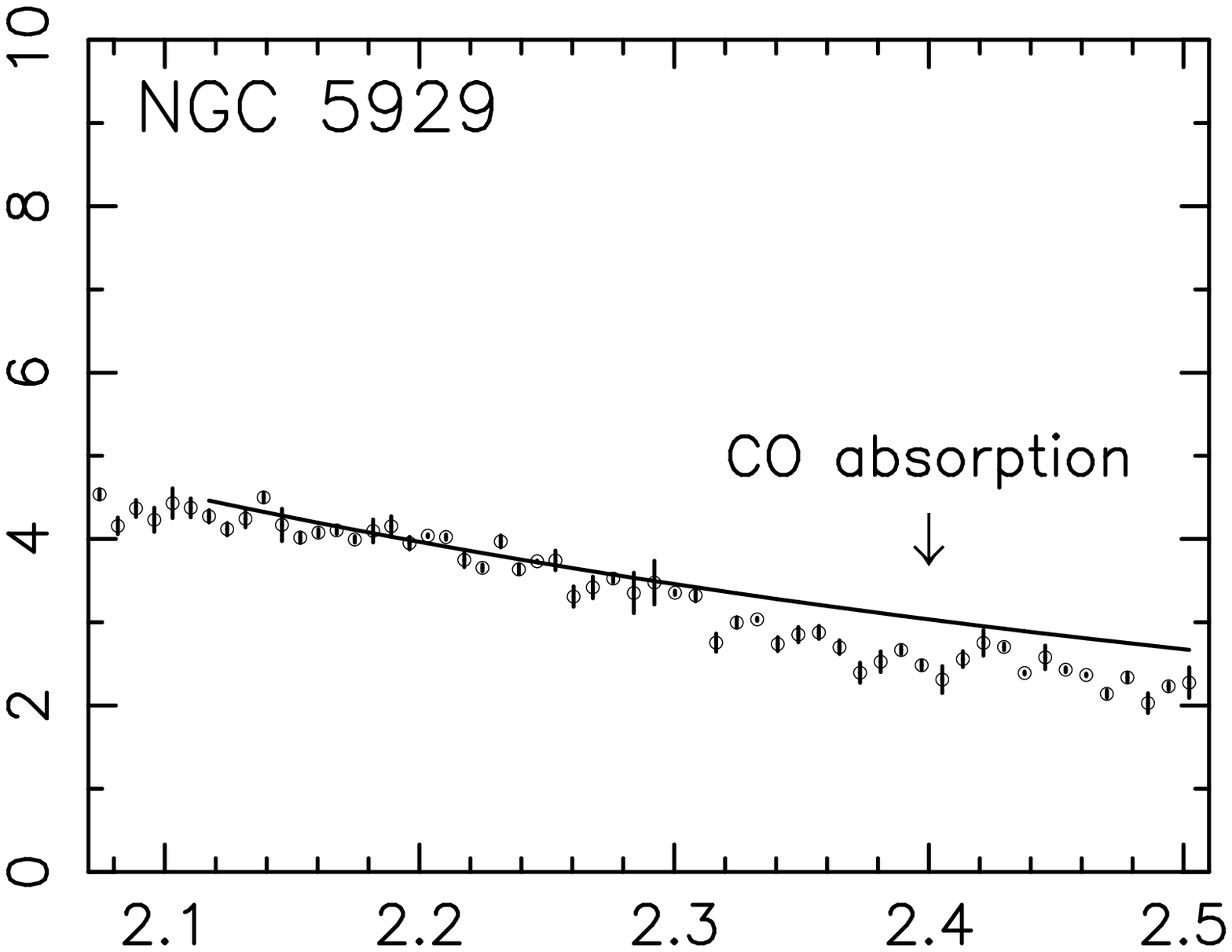} 
\includegraphics[angle=0,scale=.42]{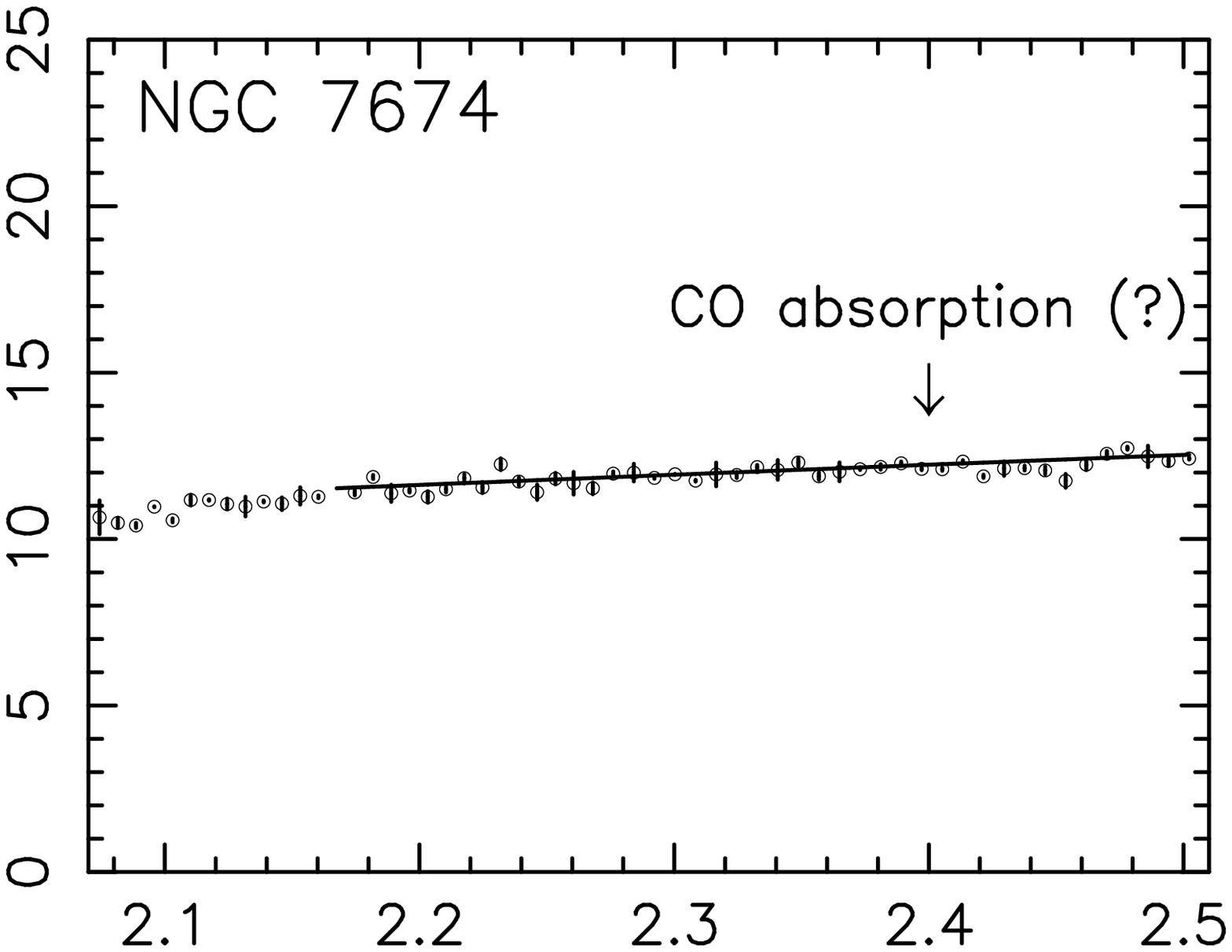} 
\includegraphics[angle=0,scale=.42]{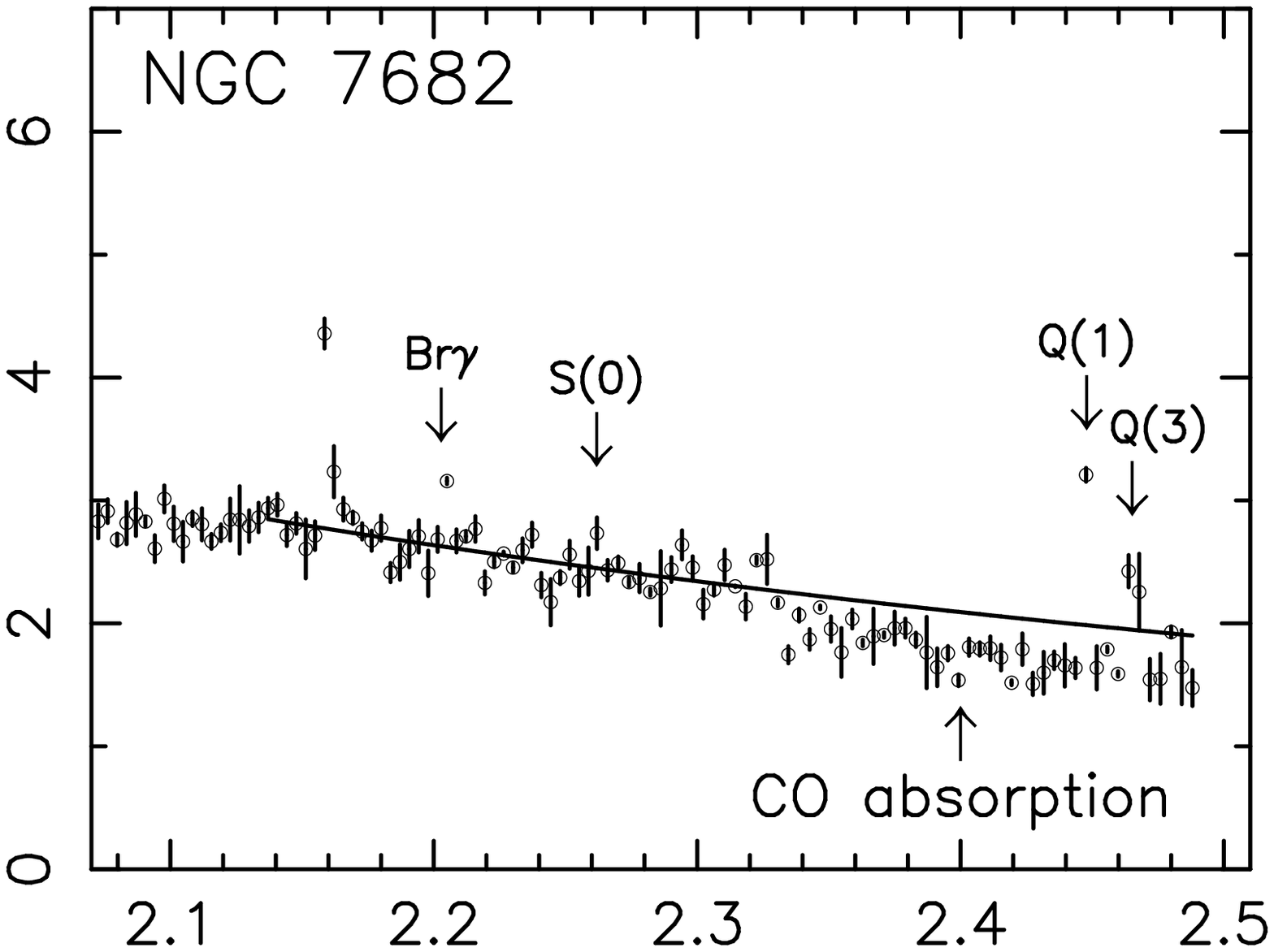} 
\includegraphics[angle=0,scale=.42]{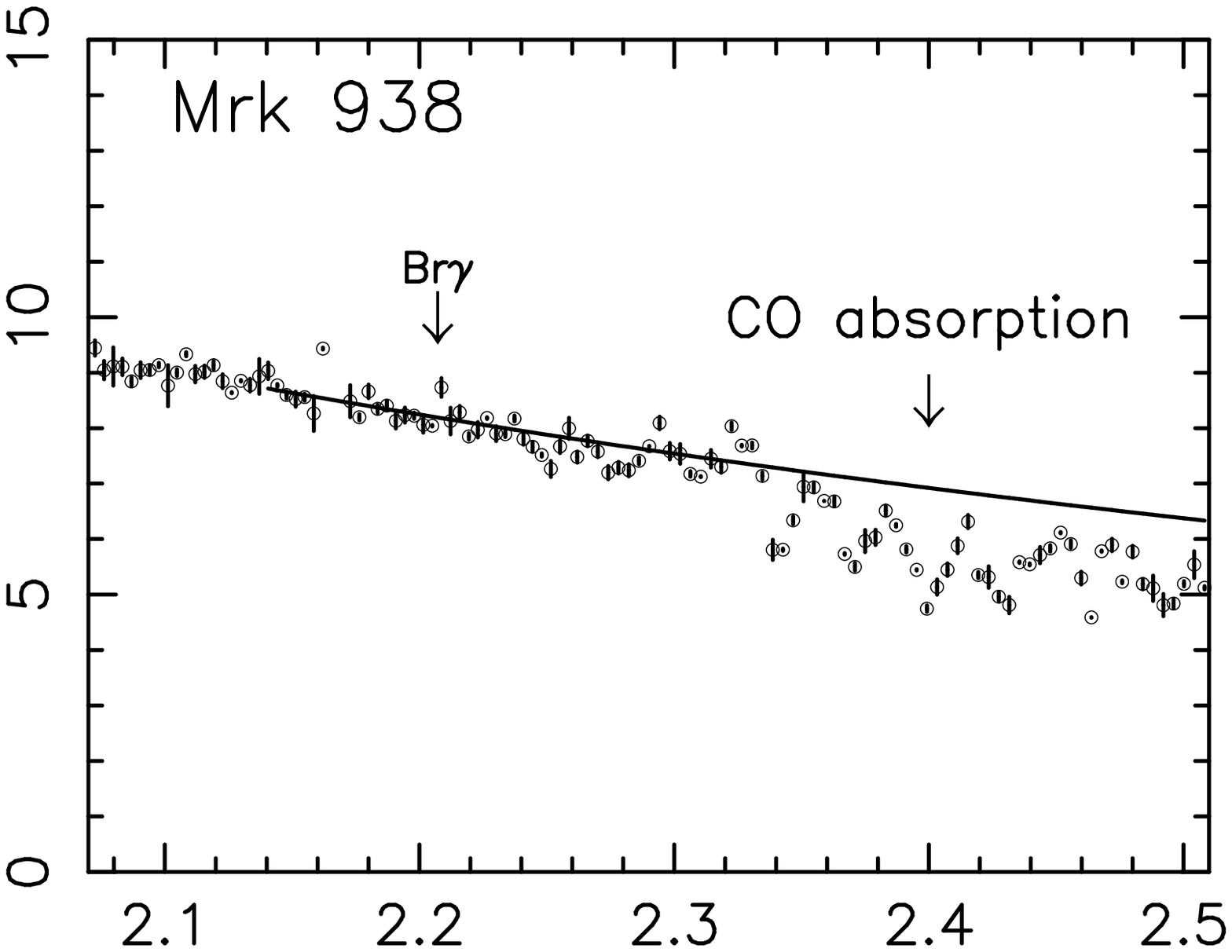} 
\includegraphics[angle=0,scale=.42]{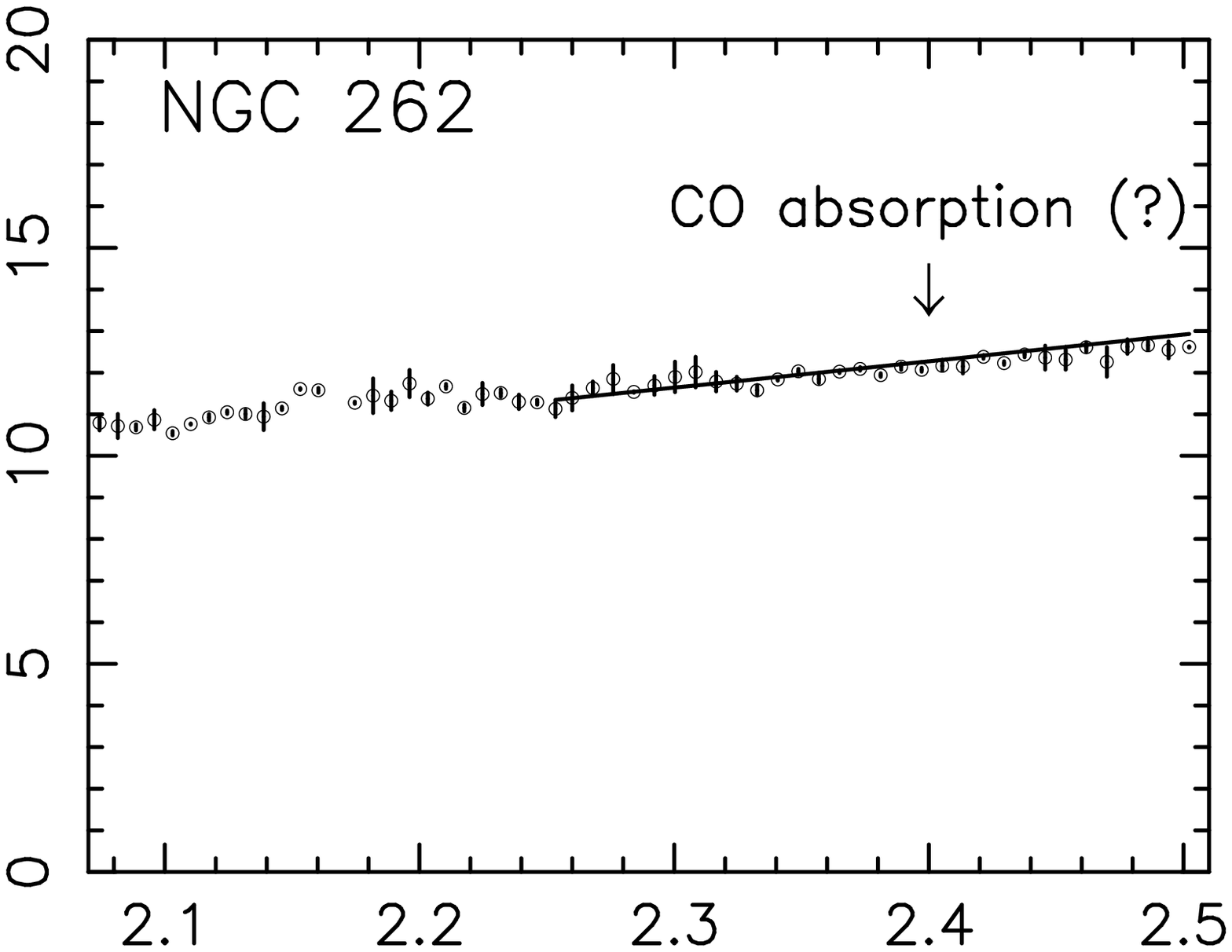} 
\end{figure}

\clearpage 

\begin{figure} 
\includegraphics[angle=0,scale=.42]{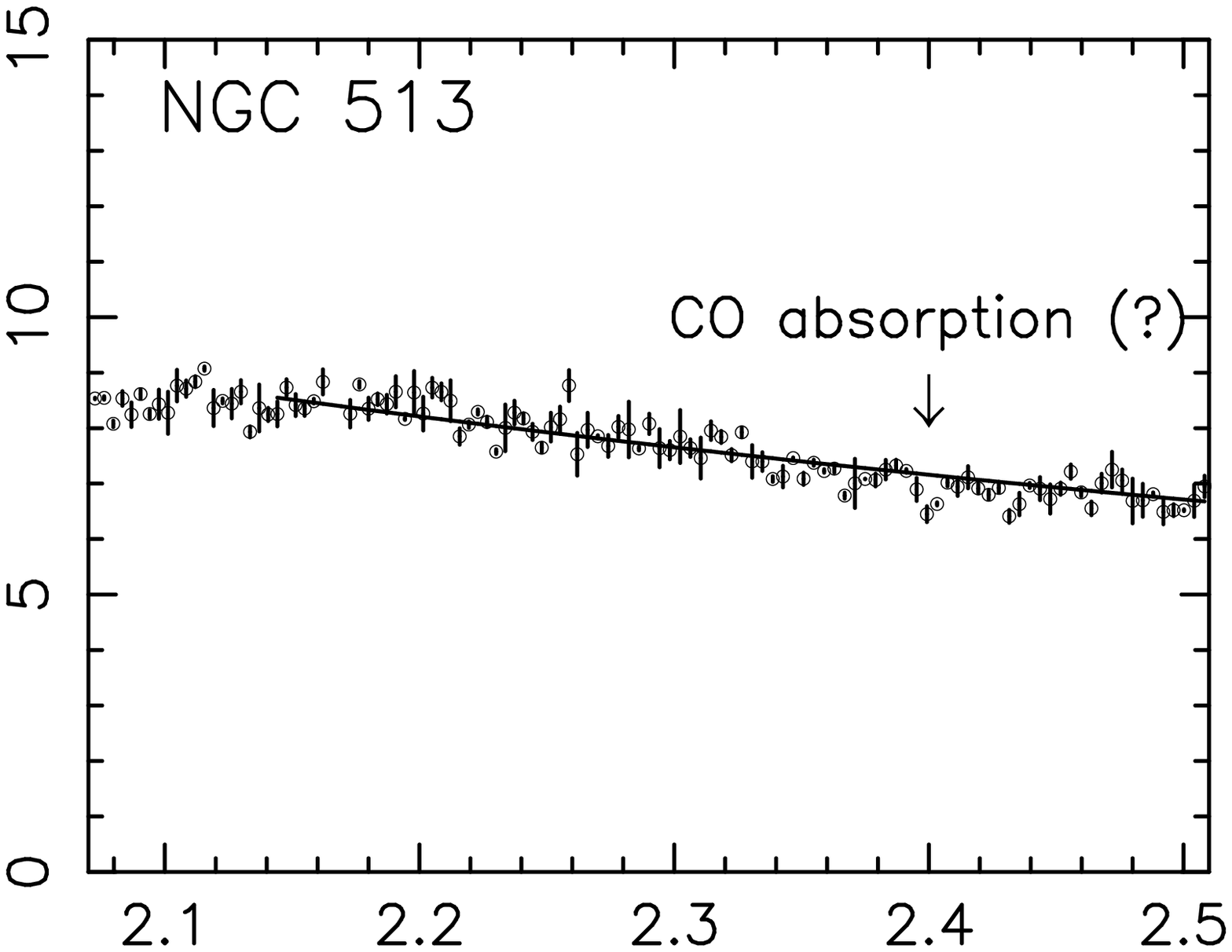} 
\includegraphics[angle=0,scale=.42]{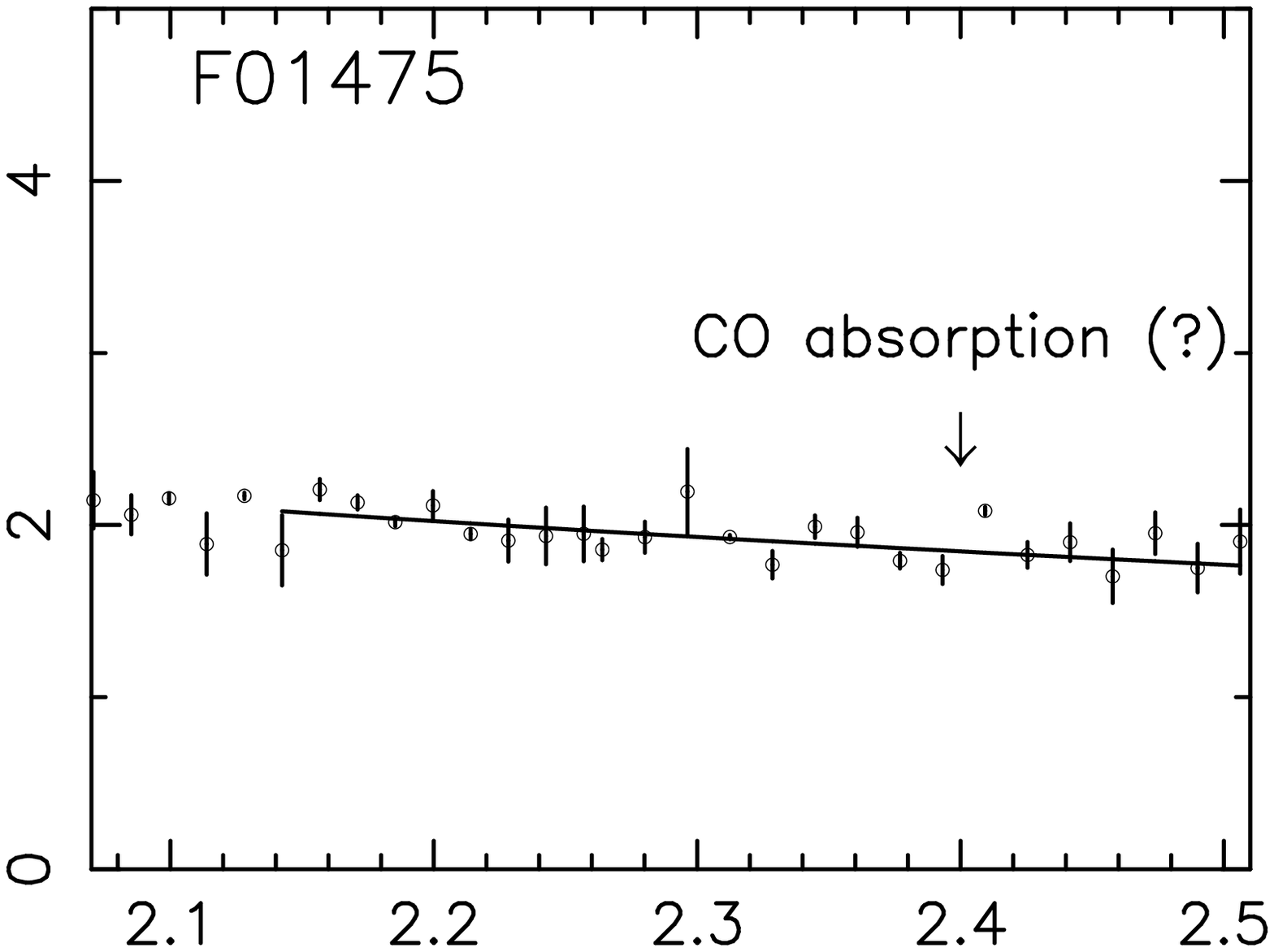} 
\includegraphics[angle=0,scale=.42]{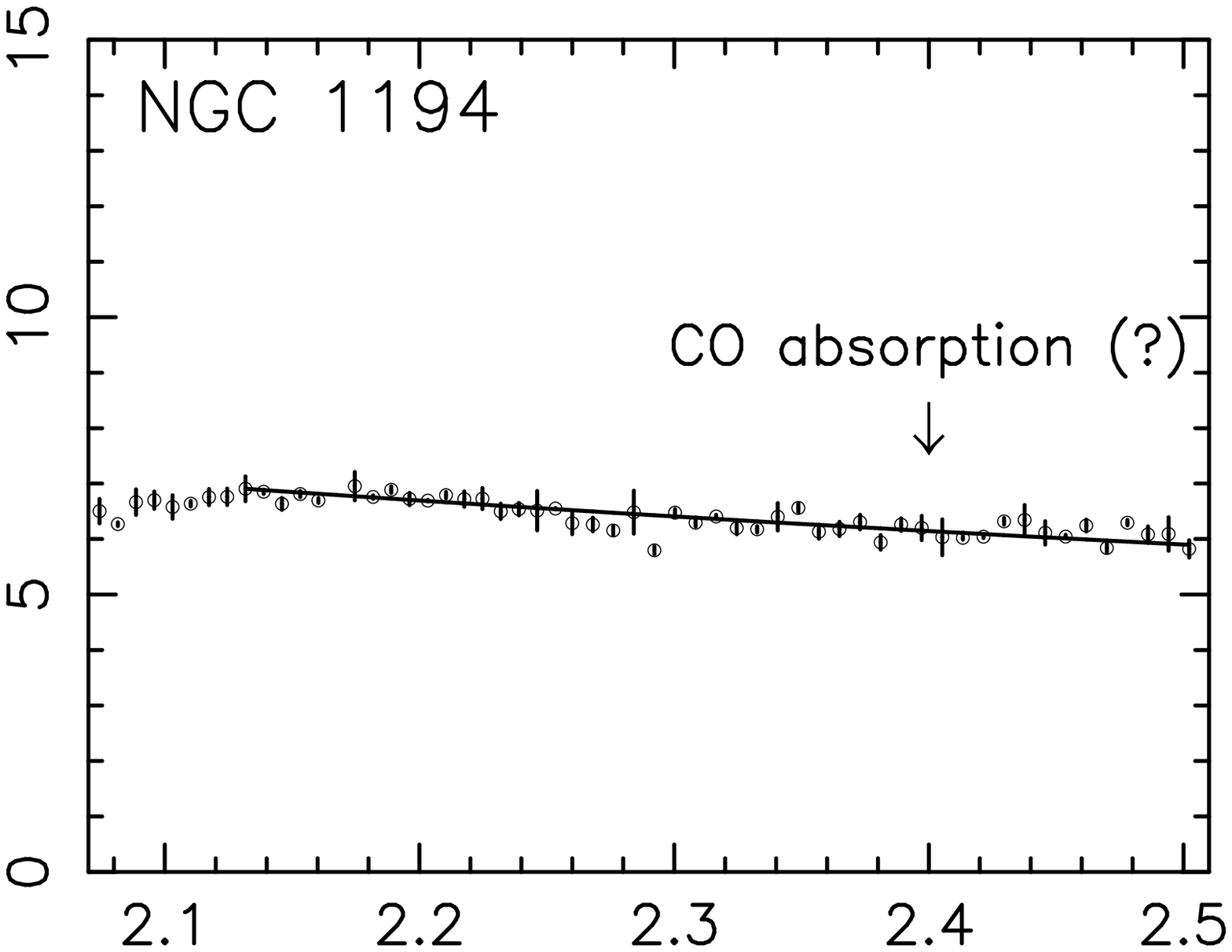} 
\includegraphics[angle=0,scale=.42]{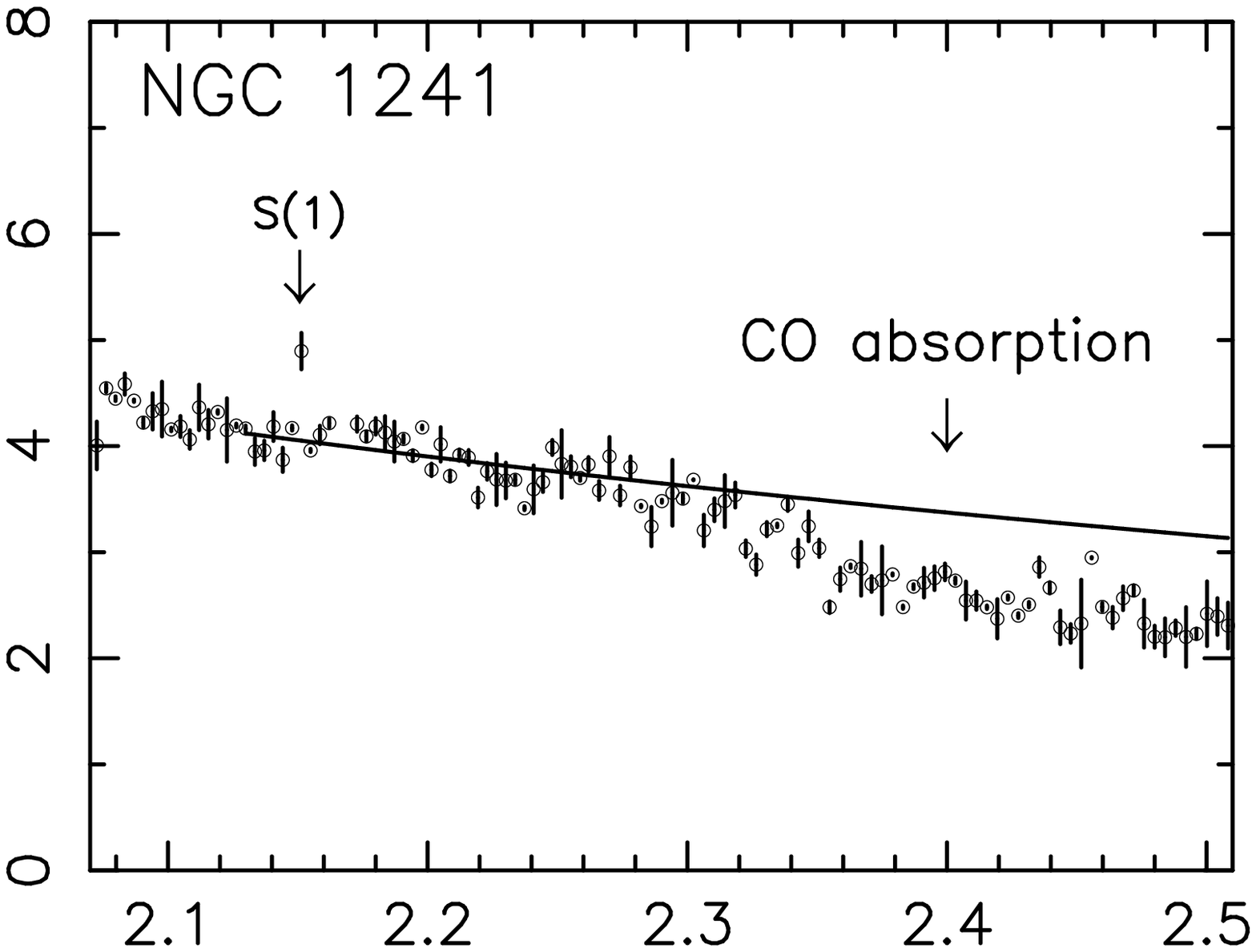} 
\includegraphics[angle=0,scale=.42]{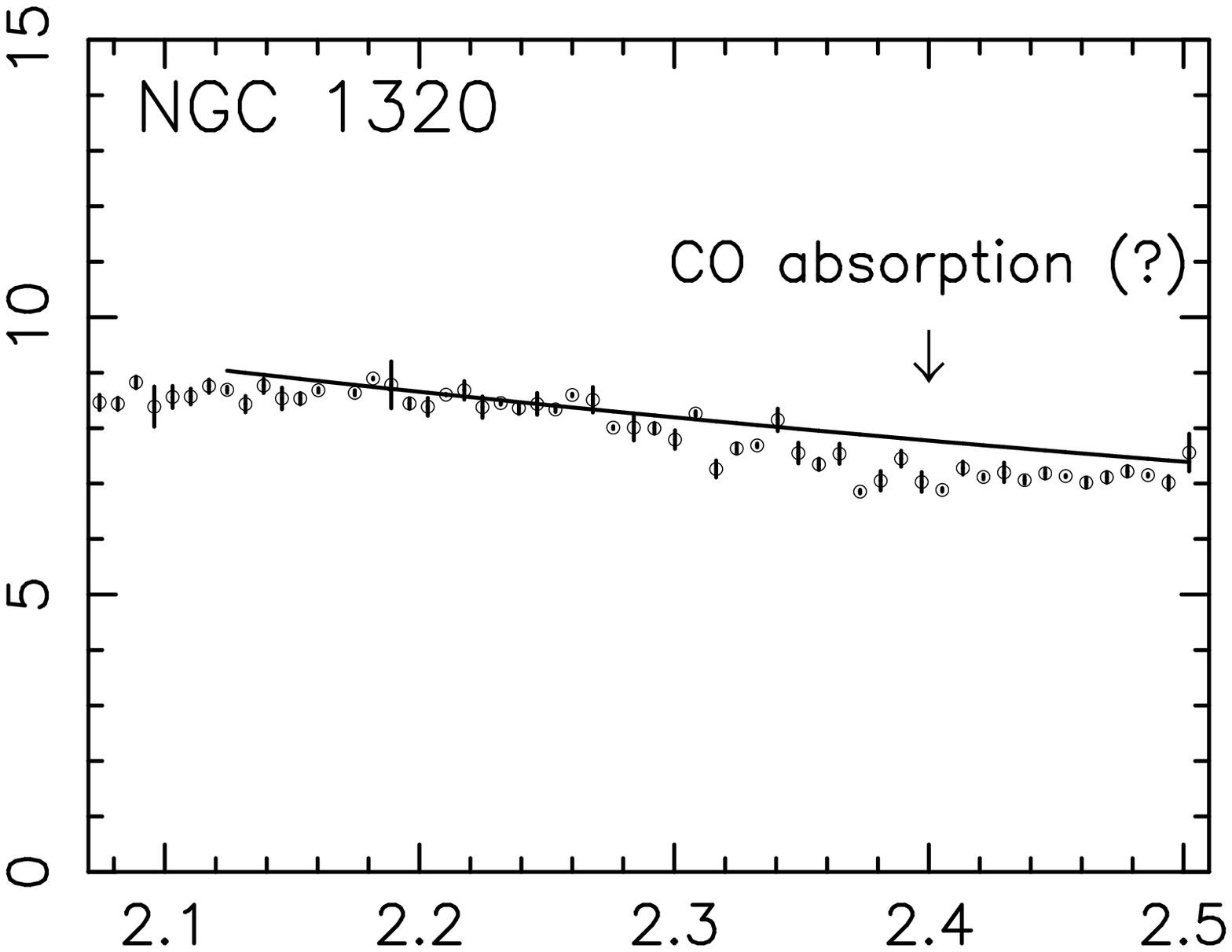} 
\includegraphics[angle=0,scale=.42]{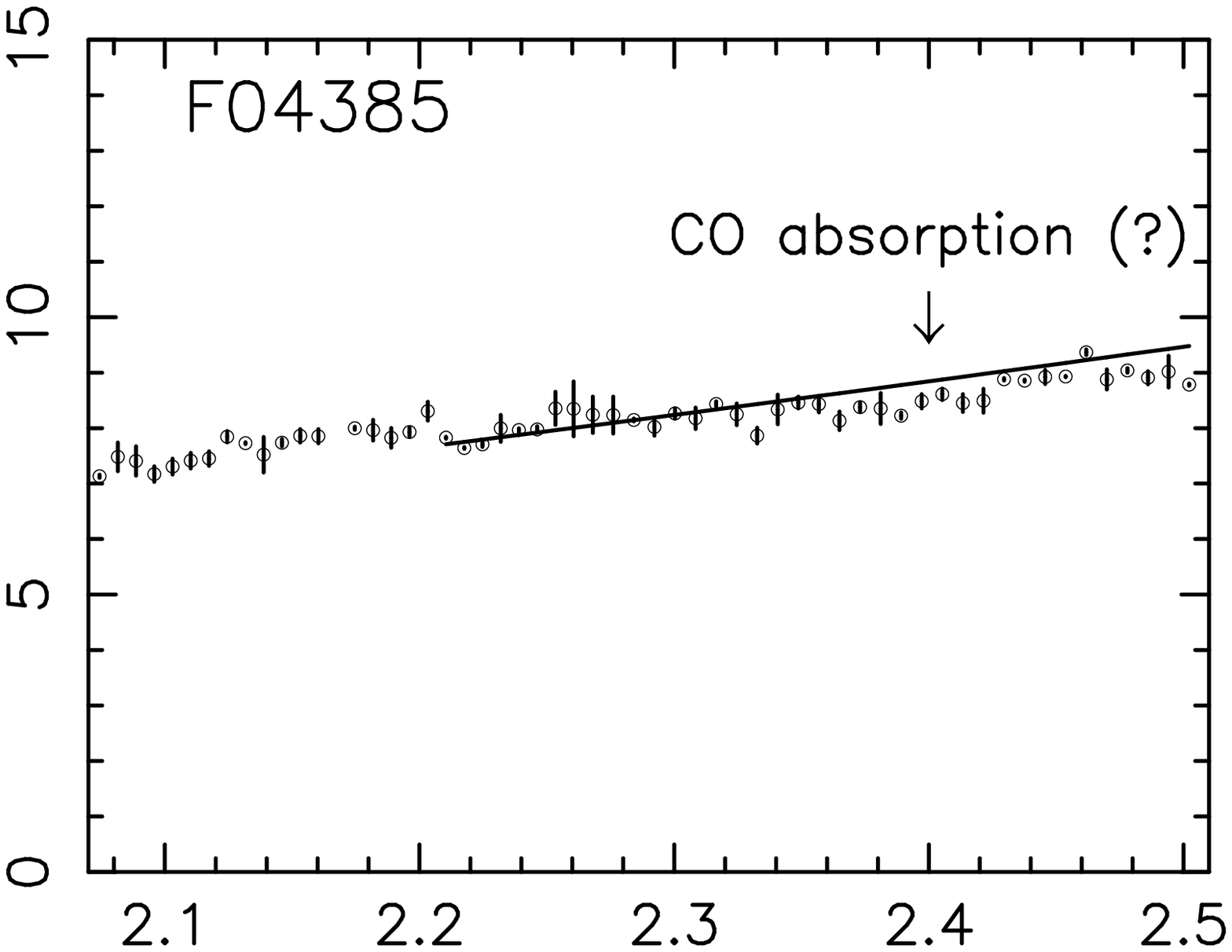} 
\end{figure}

\clearpage 

\begin{figure} 
\includegraphics[angle=0,scale=.42]{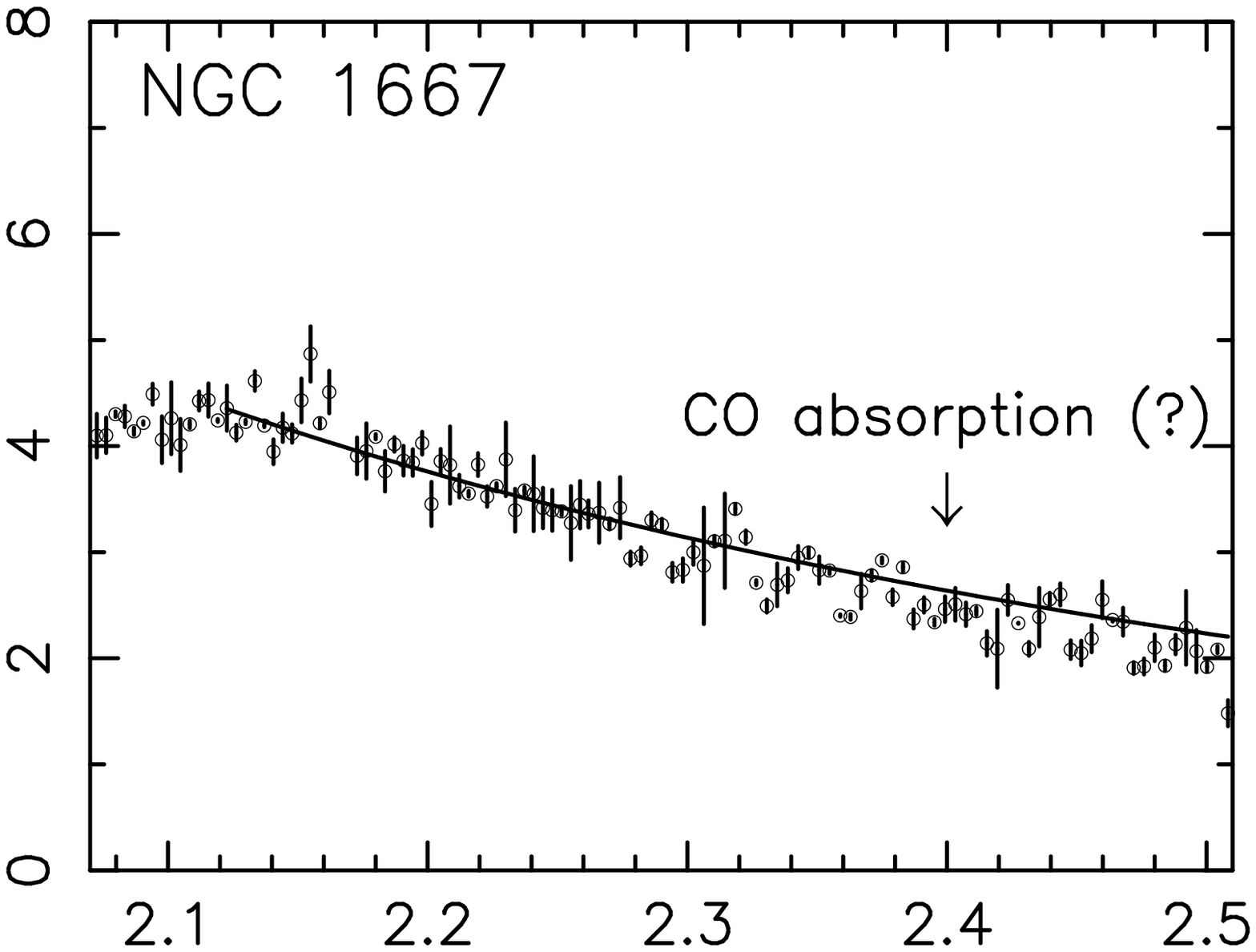} 
\includegraphics[angle=0,scale=.42]{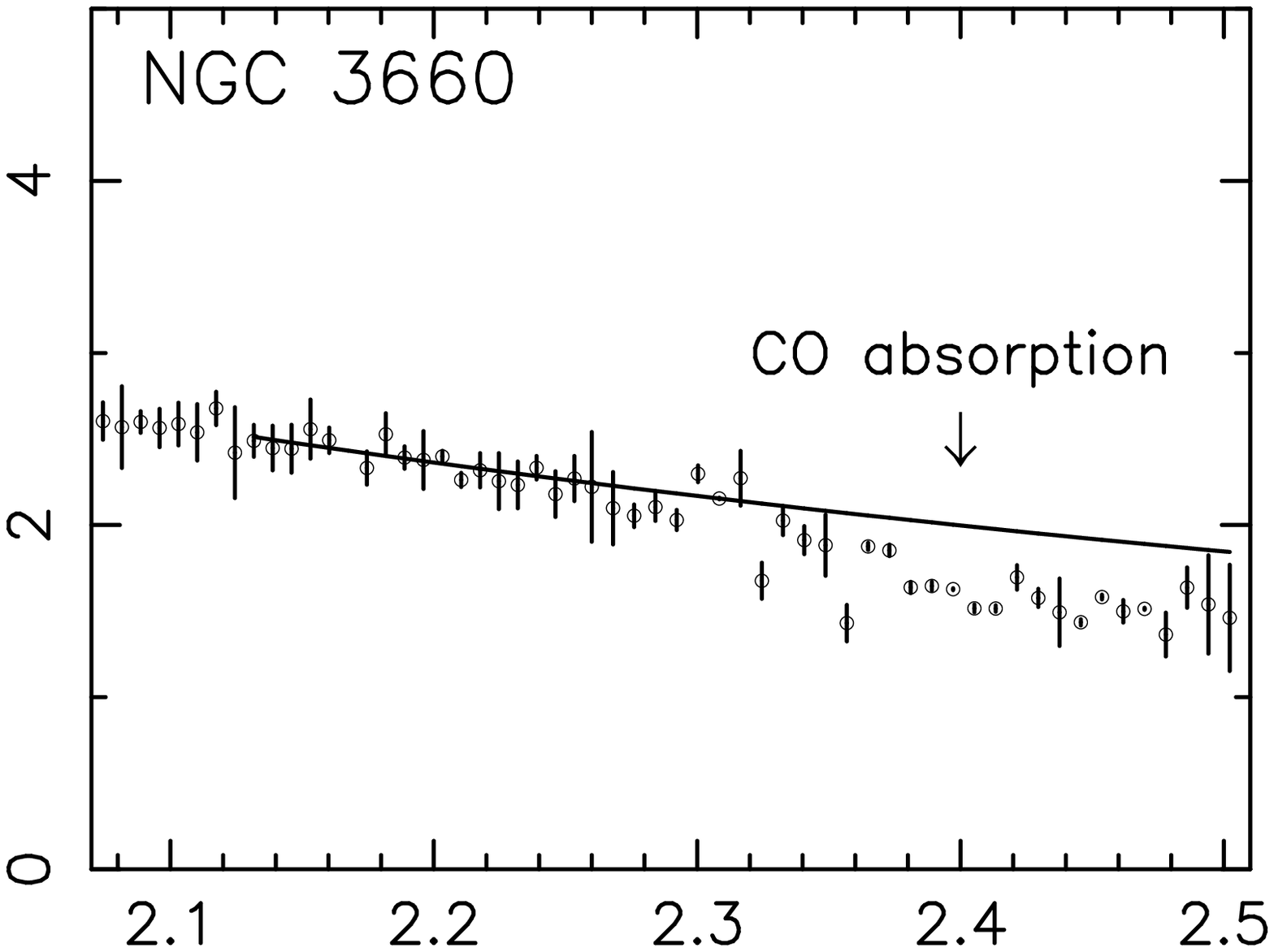} 
\includegraphics[angle=0,scale=.42]{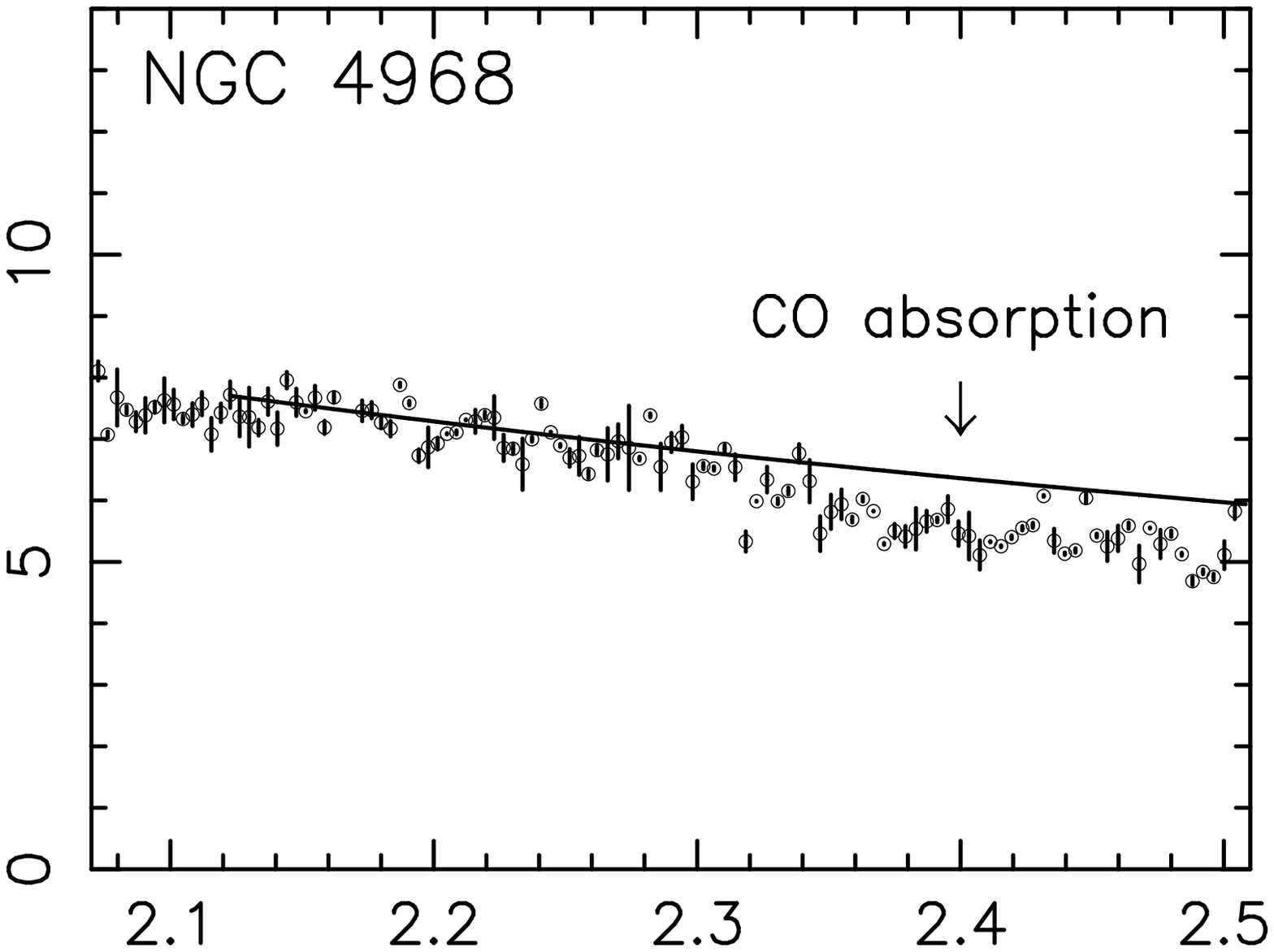} 
\includegraphics[angle=0,scale=.42]{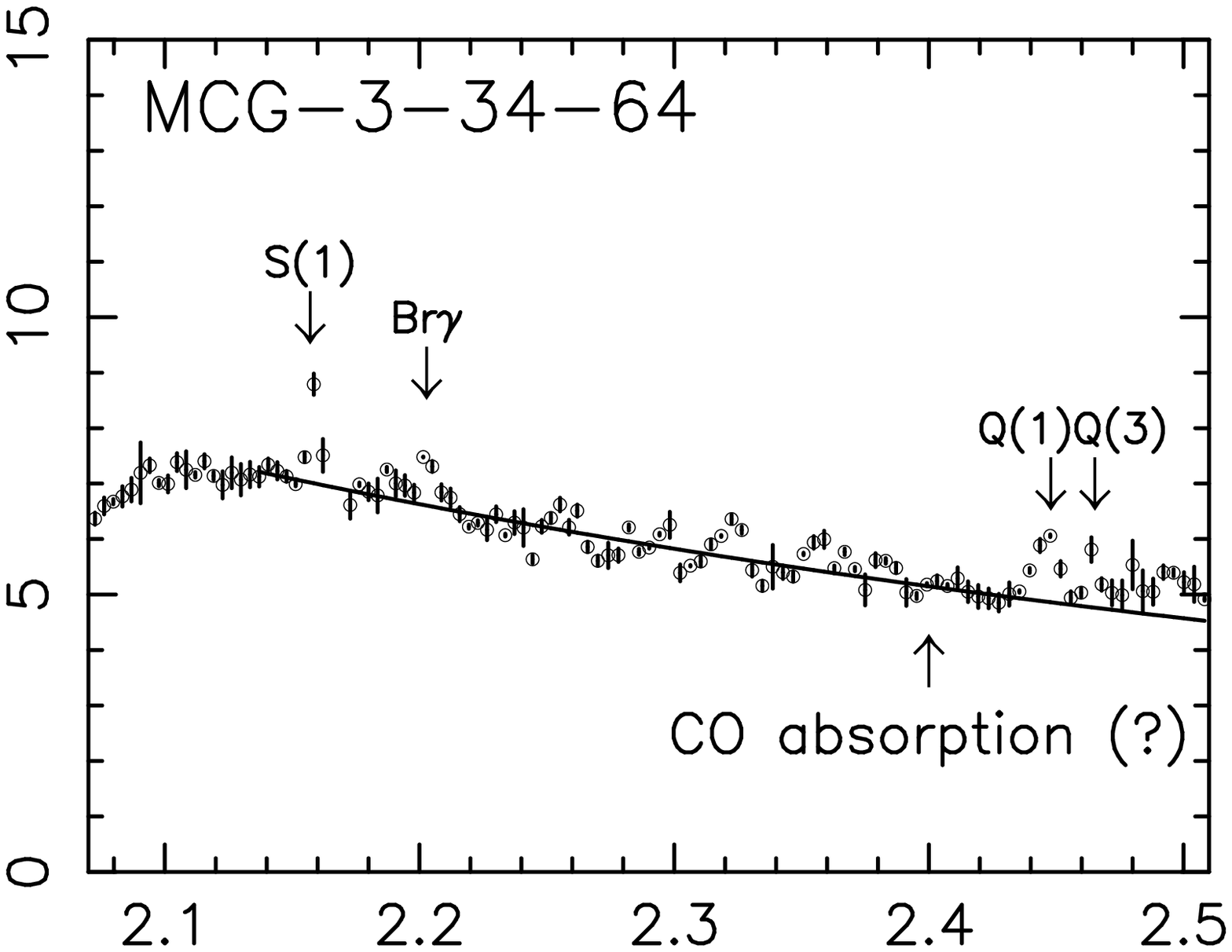} 
\includegraphics[angle=0,scale=.42]{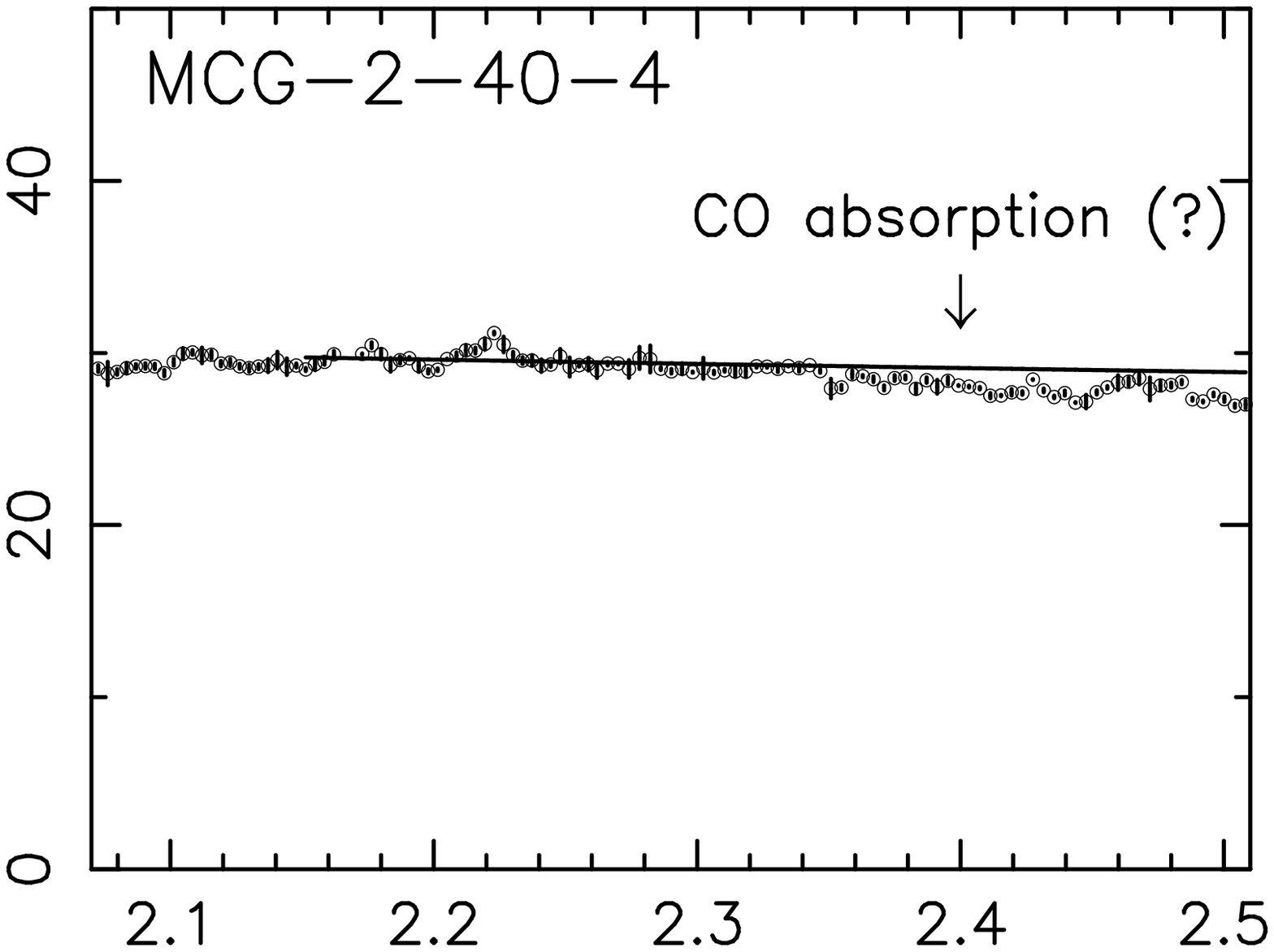} 
\includegraphics[angle=0,scale=.42]{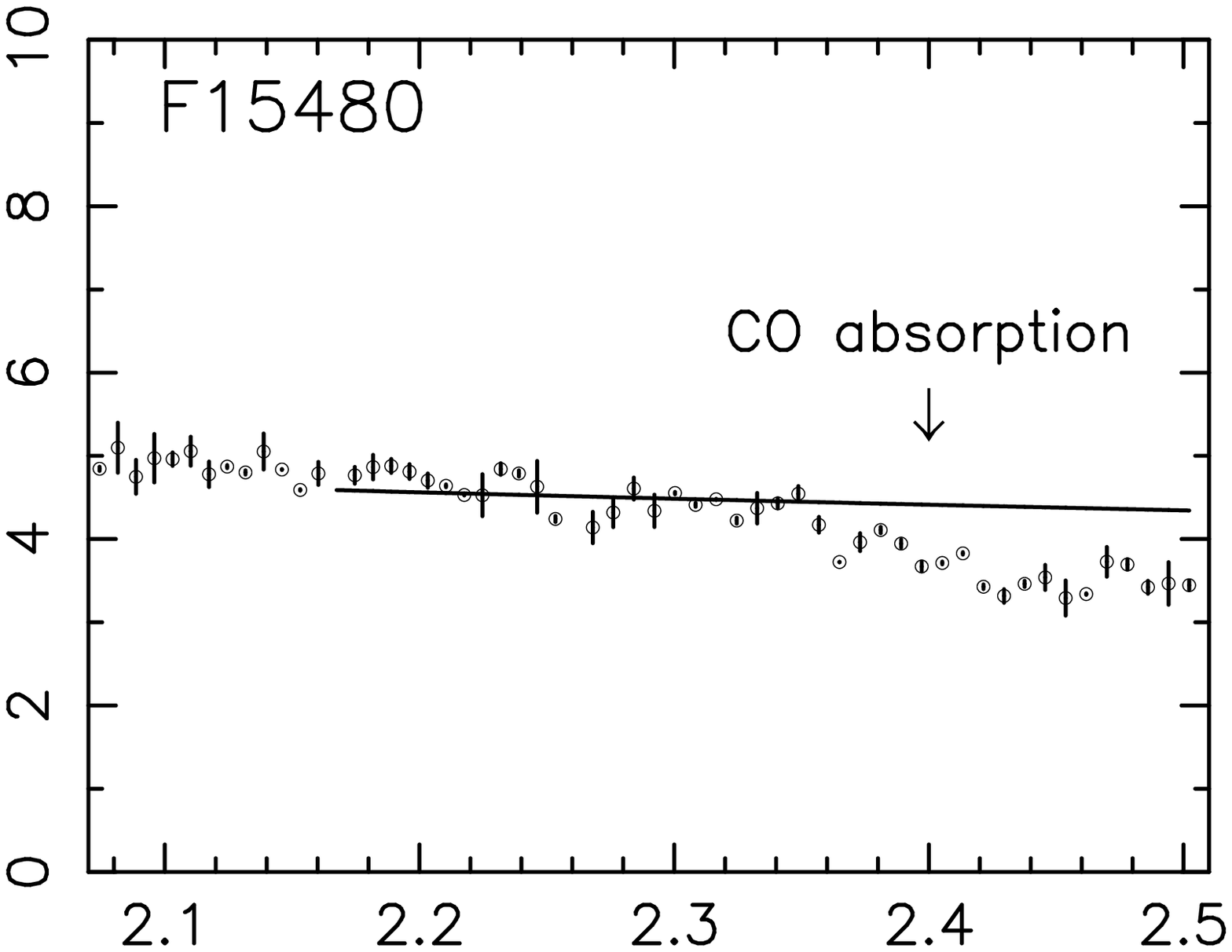} 
\end{figure}

\clearpage 

\begin{figure} 
\includegraphics[angle=0,scale=.42]{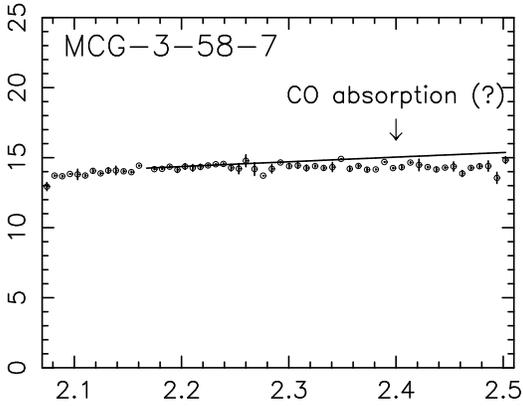} 
\caption{Infrared 2.07--2.5 $\mu$m spectra of the 25 Seyfert 2 nuclei.
The abscissa and ordinate are the observed wavelength in $\mu$m and
flux F$_{\lambda}$ in 10$^{-15}$ W m$^{-2}$ $\mu$m$^{-1}$, respectively.
The solid lines are the adopted continuum levels, with respect to which
the CO indices are measured. 
The arrows indicate the CO absorption features. 
When an observed CO index is $>$ 0.1 ($<$ 0.1), the term ``CO
absorption'' [``CO absorption (?)''] is used. 
Some strong emission lines are indicated; 
H$_{2}$ 1--0 S(1) emission line at $\lambda_{\rm rest}$ = 2.122 $\mu$m 
(denoted as S(1)), 
Br$\gamma$ $\lambda_{\rm rest}$ = 2.166 $\mu$m (Br$\gamma$), 
H$_{2}$ 1--0 S(0) at $\lambda_{\rm rest}$ = 2.223 $\mu$m (S(0)), 
H$_{2}$ 1--0 Q(1) at $\lambda_{\rm rest}$ = 2.407 $\mu$m (Q(1)), and 
H$_{2}$ 1--0 Q(3) at $\lambda_{\rm rest}$ = 2.424 $\mu$m (Q(3)).
} 
\label{fig1}
\end{figure}

\clearpage

\begin{figure} 
\includegraphics[angle=0,scale=.42]{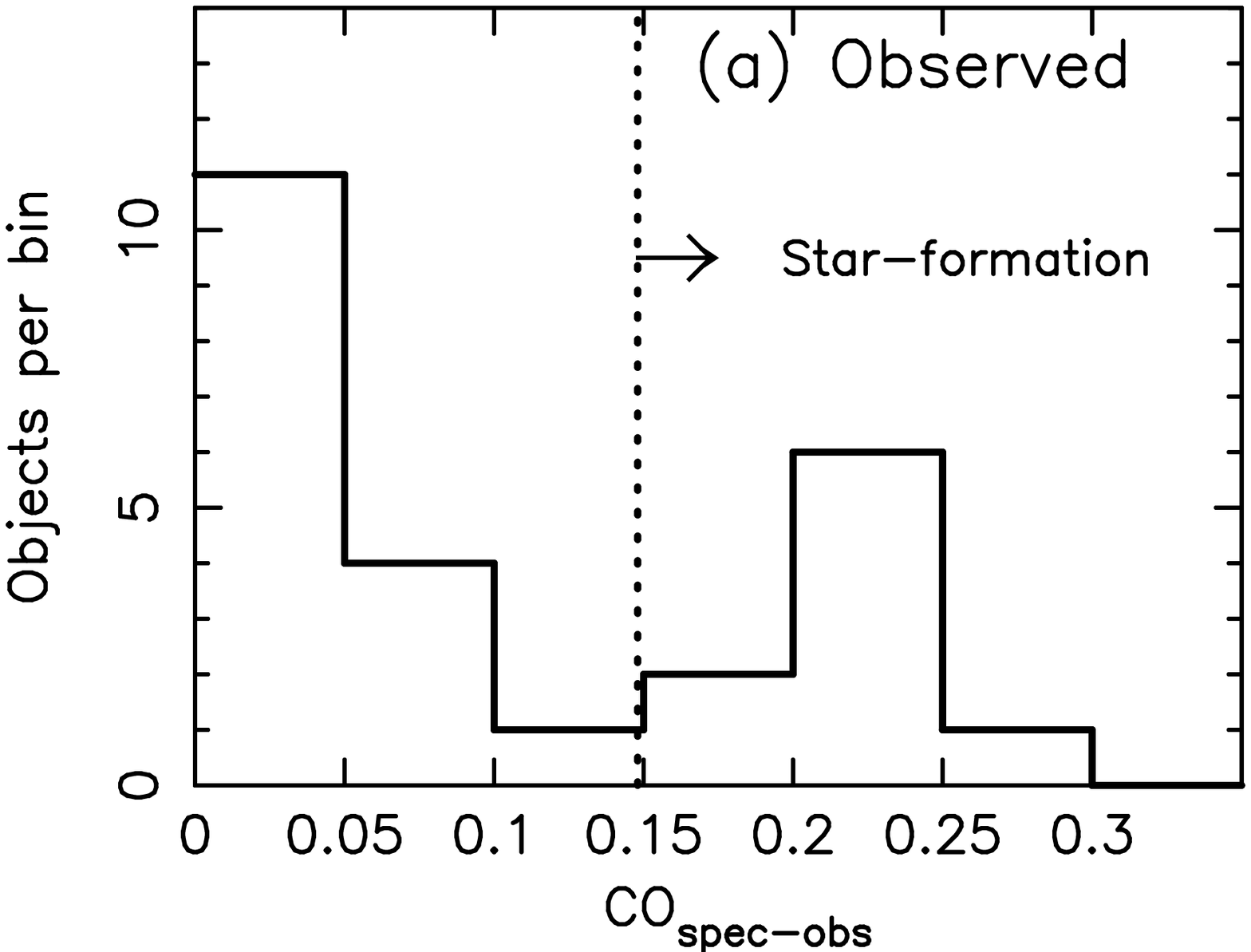} 
\includegraphics[angle=0,scale=.42]{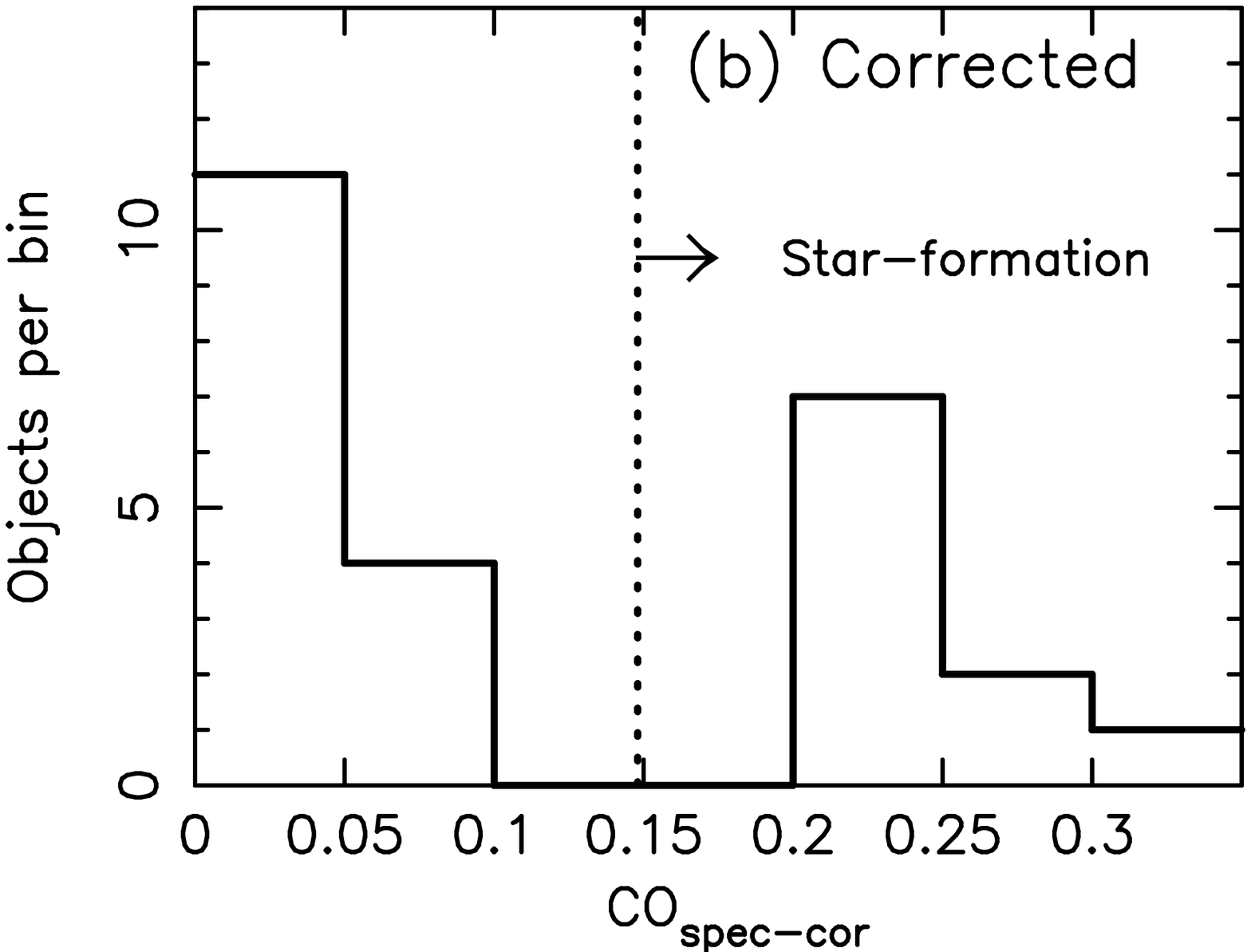} 
\caption{
{\it (a):} Histogram of the observed CO index 
(CO$_{\rm spec-obs}$).  
{\it (b):} Histogram of the corrected CO index 
(CO$_{\rm spec-cor}$).  
For sources with observed CO indices $<$ 0.10, no AGN correction was
attempted. 
Star-forming or elliptical galaxies typically show CO$_{\rm spec}$ 
$>$ 0.15 \citep{gol97a,gol97b,iva00}. 
} 
\label{fig2}
\end{figure}

\clearpage

\begin{figure} 
\includegraphics[angle=0,scale=.42]{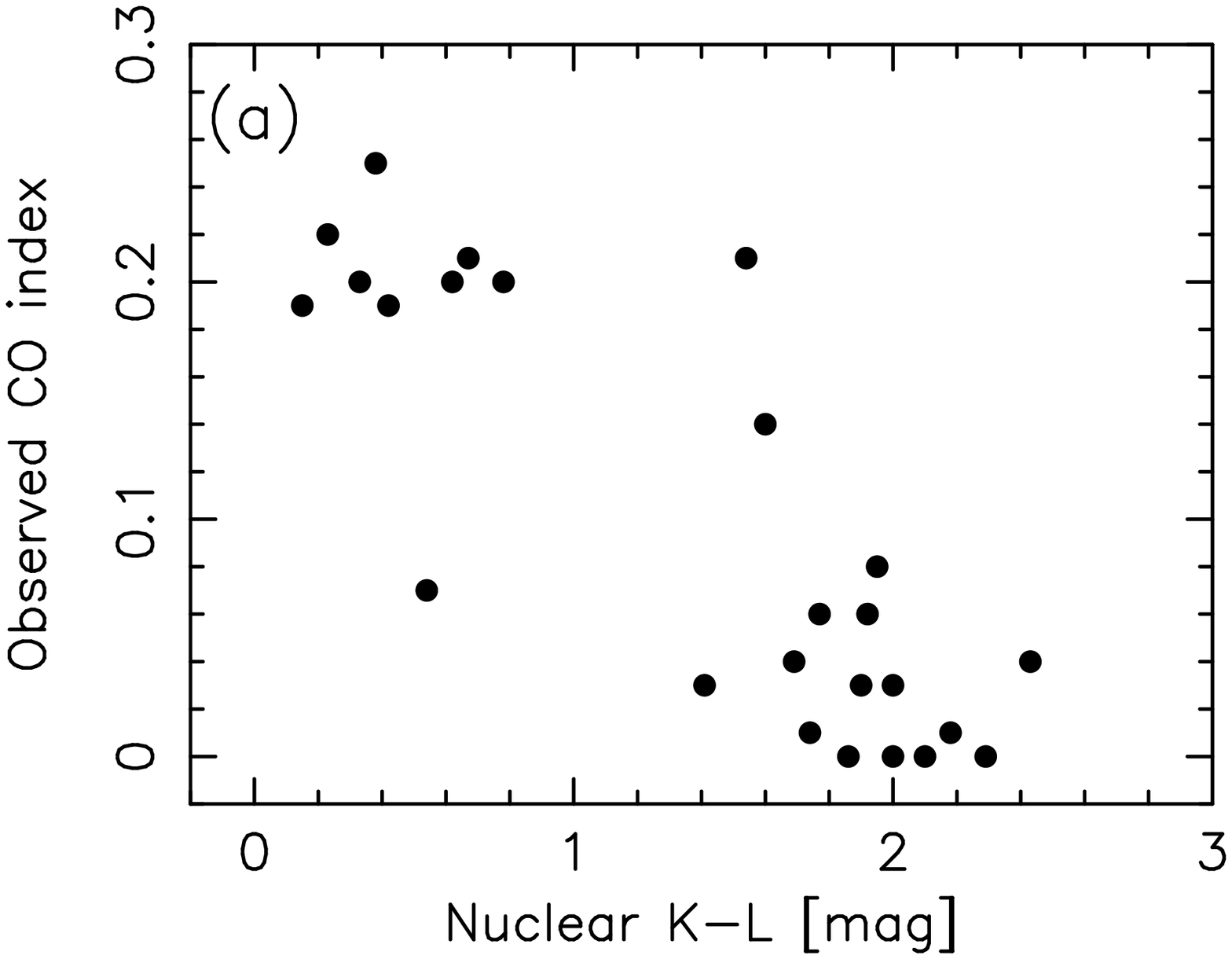} 
\includegraphics[angle=0,scale=.42]{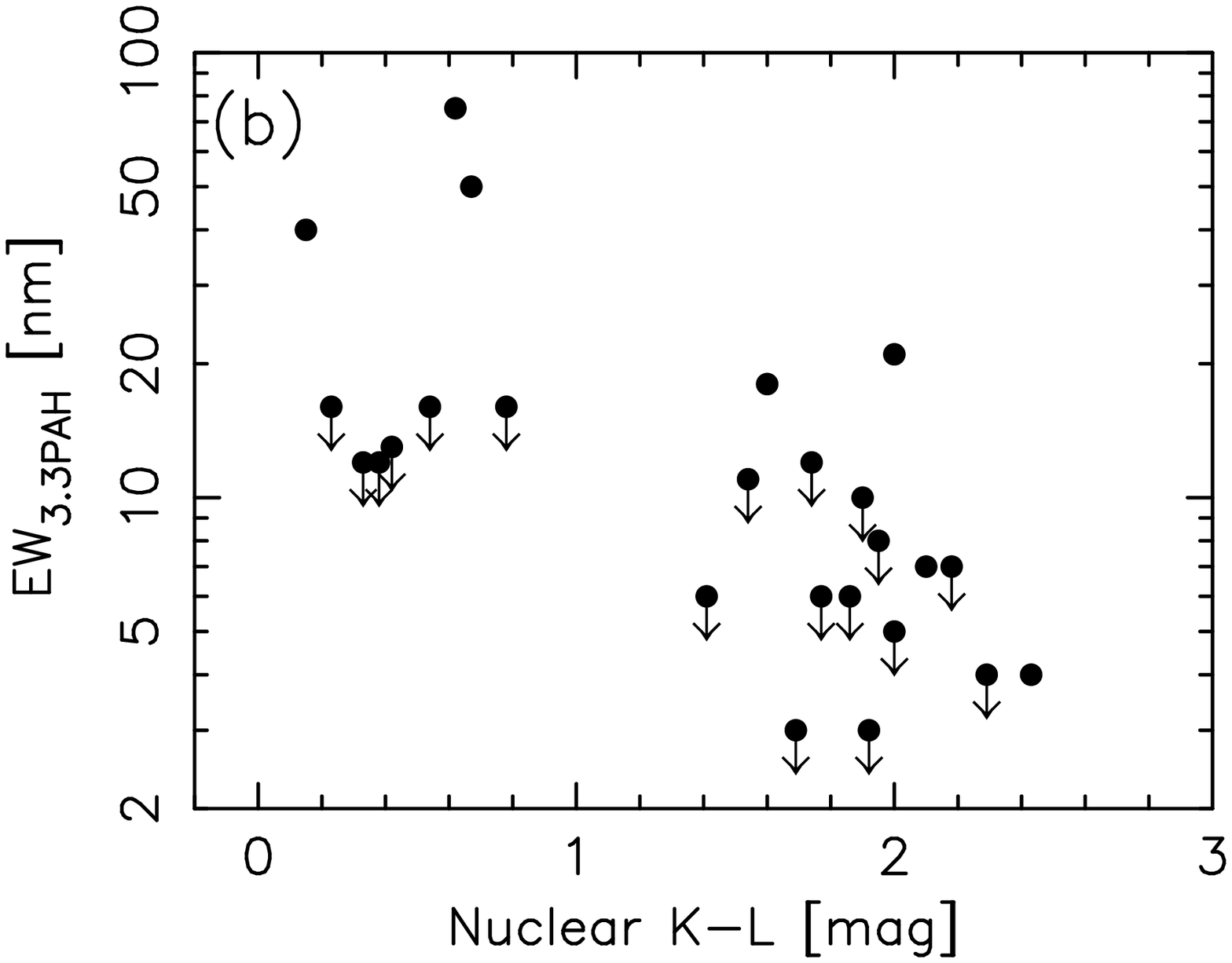} 
\caption{
{\it (a)}: Nuclear $K-L$ magnitude (abscissa) and observed spectroscopic
CO index (ordinate). 
The uncertainty of the $K-L$ color is $<$0.1 mag. 
{\it (b)}: The ordinate is the rest-frame equivalent width of
the 3.3 $\mu$m PAH emission feature. 
Its uncertainty is at a level of $<$30\% for the PAH-detected sources. 
} 
\label{fig3}
\end{figure}

\clearpage

\begin{figure} 
\includegraphics[angle=0,scale=.6]{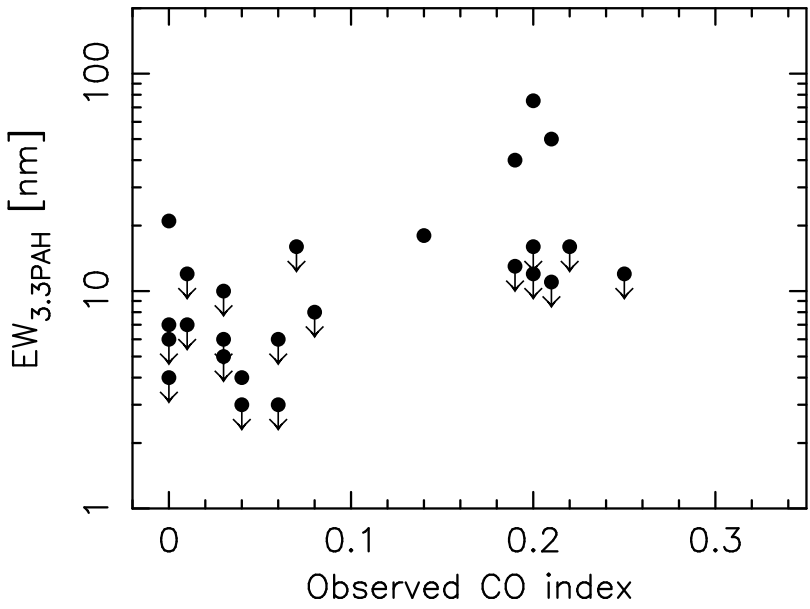} 
\caption{
Observed spectroscopic CO index (abscissa) and rest-frame
equivalent width of the 3.3 $\mu$m PAH emission feature (ordinate).
}
\label{fig4}
\end{figure}

\clearpage

\begin{figure} 
\includegraphics[angle=0,scale=.6]{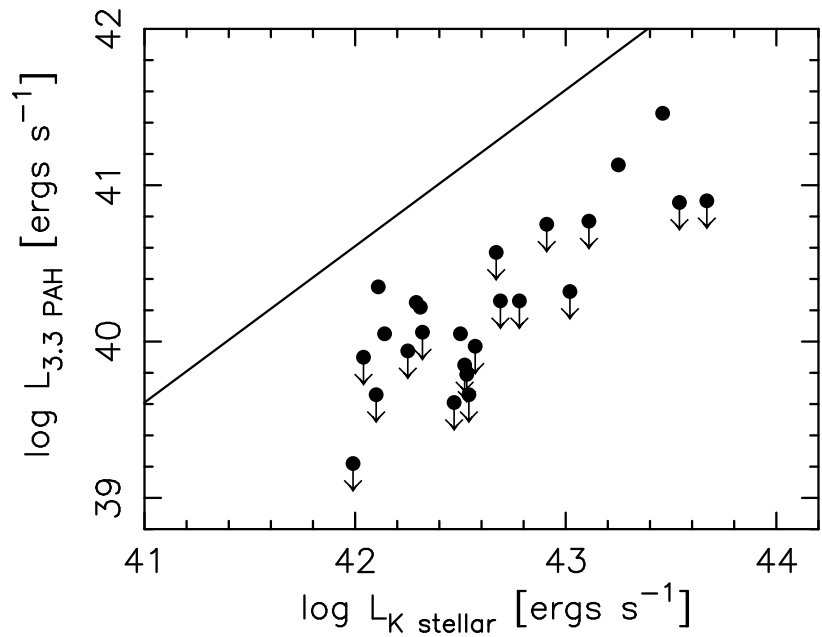} 
\caption{
The observed nuclear stellar $K$-band luminosity, defined as
$\lambda$L$_{\lambda}$ at 2.2 $\mu$m (abscissa) and nuclear 
3.3 $\mu$m PAH luminosity estimated from our slit spectra (ordinate). 
The errors of the PAH luminosities for the PAH-detected sources are at
a level of $<$30\%.  
The solid line indicates the correlation for infrared-luminous
starburst galaxies (see text).
} 
\label{fig5}
\end{figure}


\clearpage

\begin{figure} 
\includegraphics[angle=0,scale=.42]{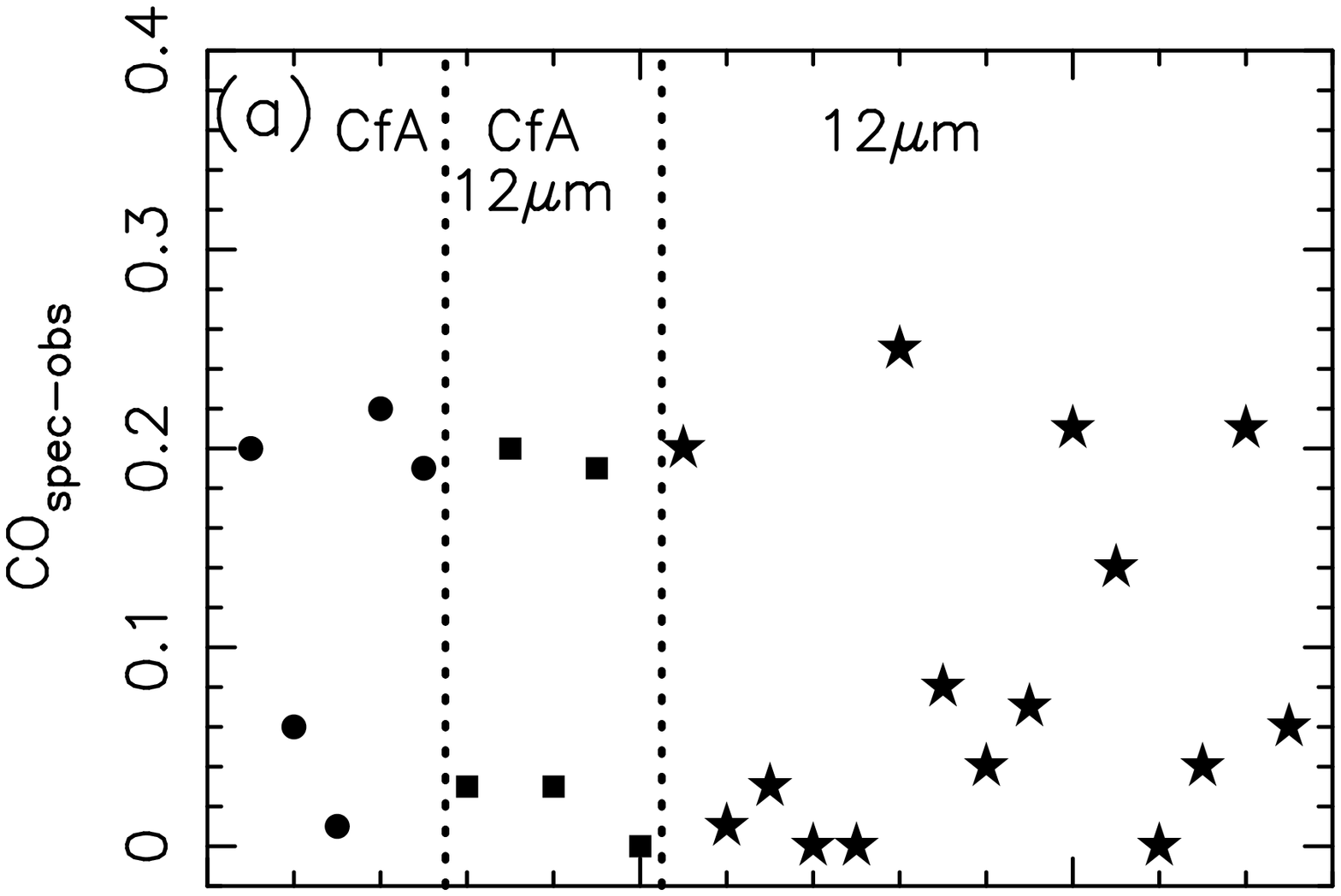} 
\includegraphics[angle=0,scale=.42]{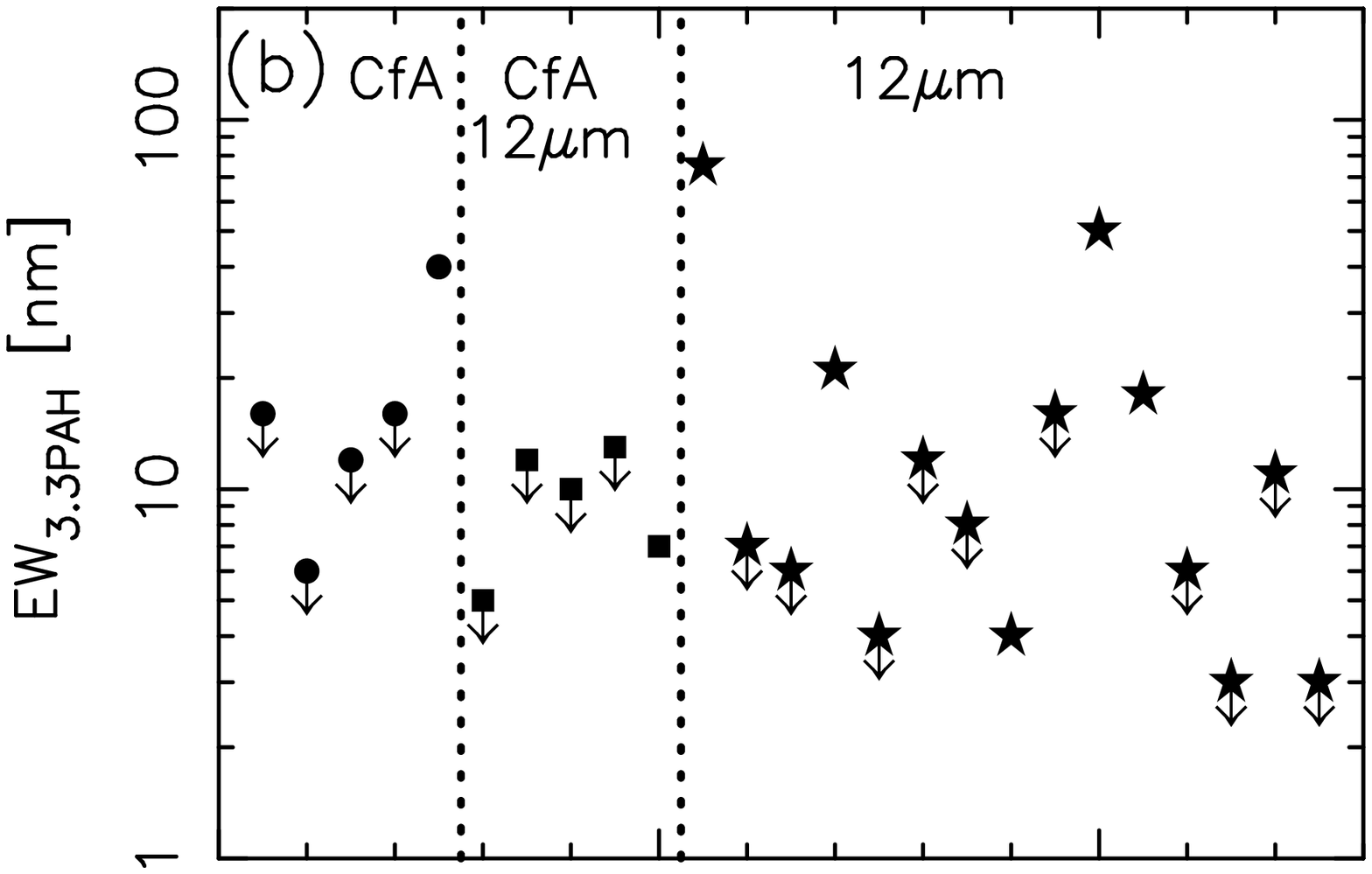} 
\includegraphics[angle=0,scale=.42]{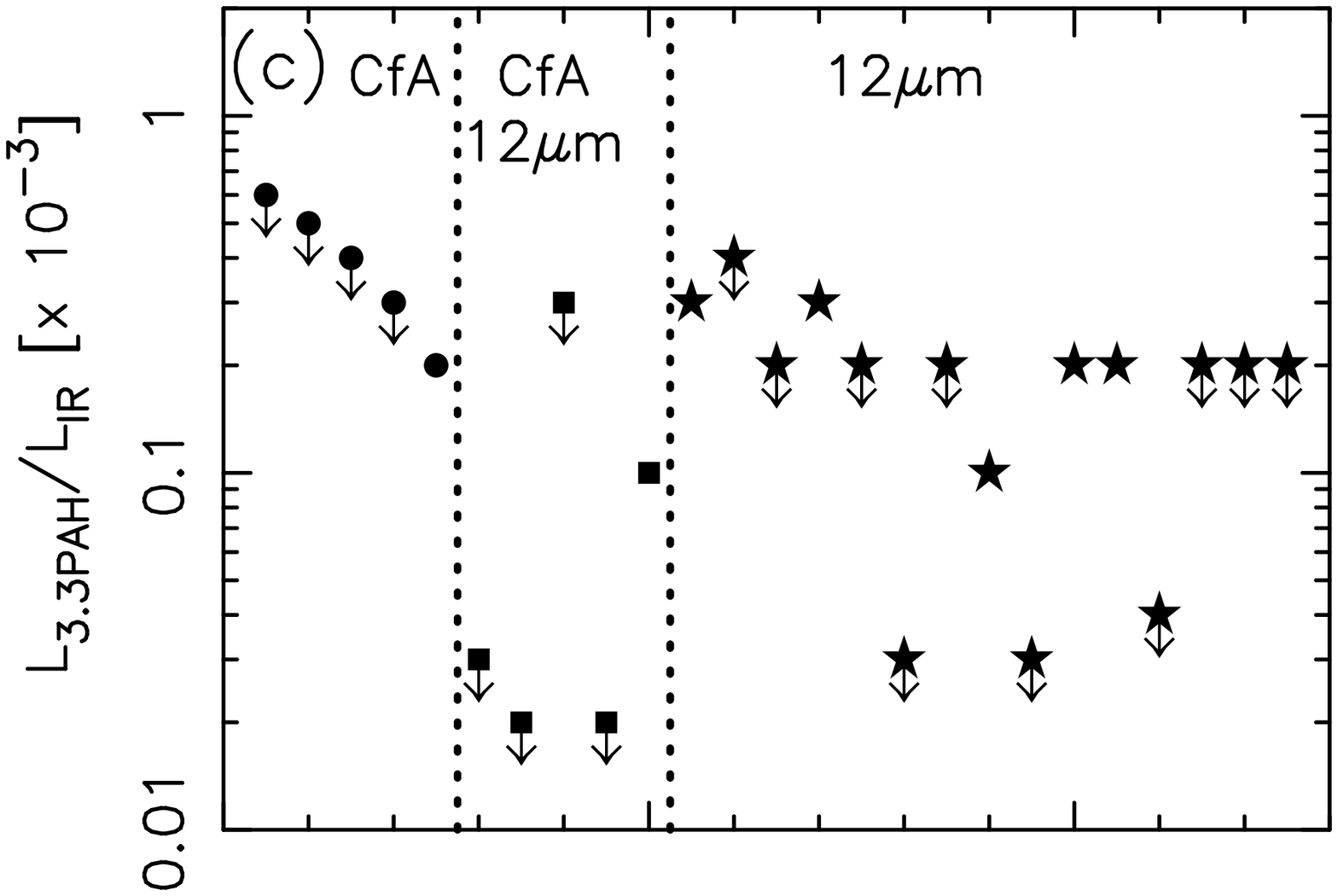} 
\includegraphics[angle=0,scale=.42]{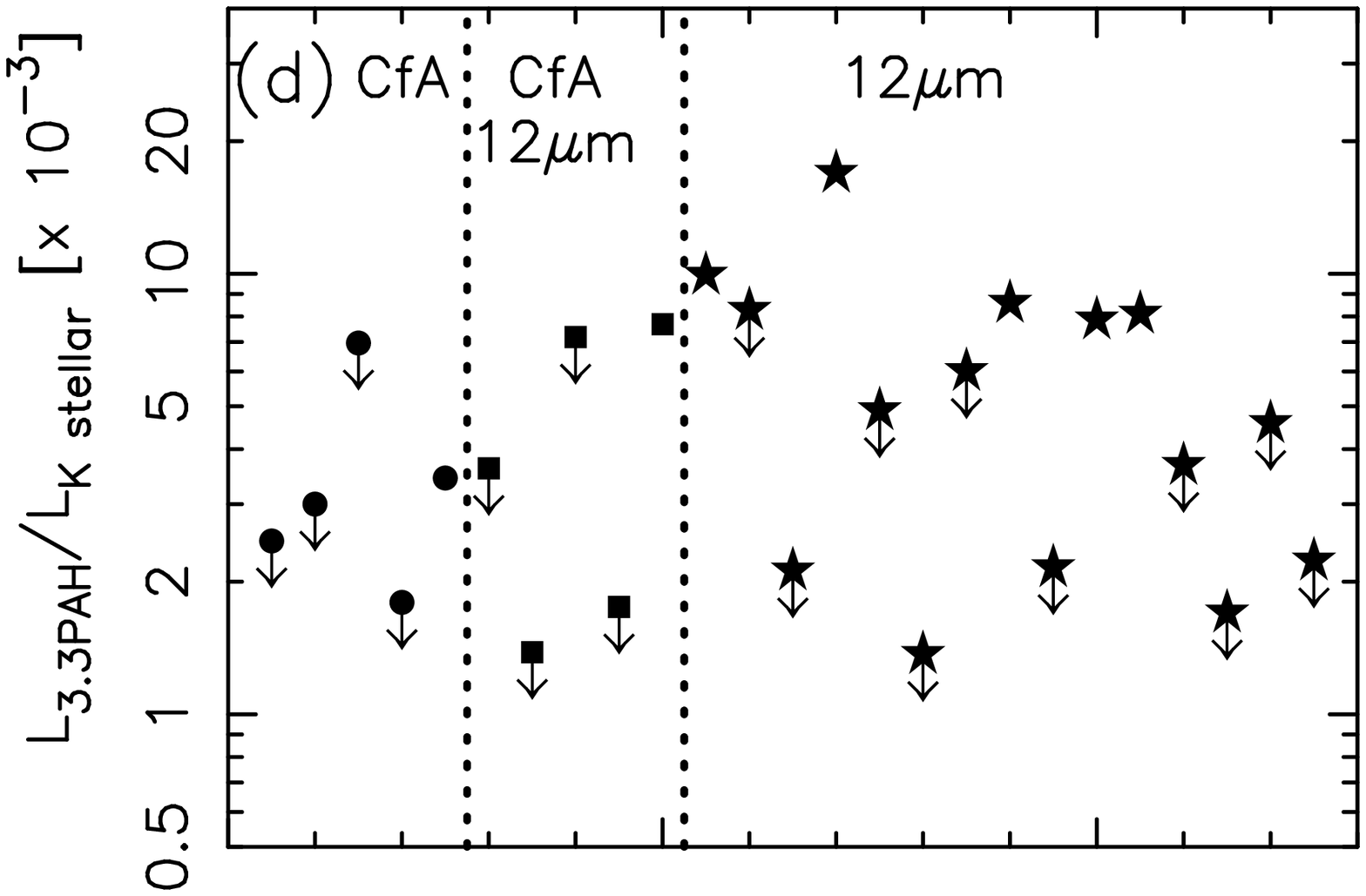}
\caption{
{\it (a)}: The distribution of the observed spectroscopic CO index.
Solid circles: CfA Seyfert 2s. Solid stars: 12 $\mu$m Seyfert 2s.  
Five sources listed in both samples are plotted as solid squares. 
{\it (b)}: The rest-frame equivalent width of the 3.3 $\mu$m PAH
emission feature. 
{\it (c)}: The 3.3 $\mu$m PAH to infrared luminosity ratio, taken from
\citet{ima03}. 
{\it (d)}: The 3.3 $\mu$m PAH to stellar $K$-band luminosity ratio.
} 
\label{fig6}
\end{figure}

\end{document}